\documentstyle[twocolumn,prl,aps]{revtex} 
\begin{document} 
\draft 
\newcommand{\lapx}{\stackrel{<}{\scriptstyle \sim}} 
\newcommand{\gapx}{\stackrel{>}{\scriptstyle \sim}} 
\def\half{{1\over 2}} 
\def\m{\ell}
\def\vopo{(VO)$_2$P$_2$O$_7$\ }  
\def\cuno{Cu(NO$_3$)$_2\cdot 2.5$H$_2$O\ } 
\def\cugeons{CuGeO$_3$} 
\def\cugeo{CuGeO$_3$\ } 
\def\vopons{(VO)$_2$P$_2$O$_7$}  
\def\cunons{Cu(NO$_3$)$_2\cdot 2.5$H$_2$O} 
\def\navo{$\alpha'$-NaV$_2$O$_5$\ } 
\def\navons{$\alpha'$-NaV$_2$O$_5$} 
\newcommand{\lapprox}{\stackrel{<}{\scriptstyle \sim}} 
\newcommand{\gapprox}{\stackrel{>}{\scriptstyle \sim}}

\twocolumn 
[\hsize\textwidth\columnwidth\hsize\csname @twocolumnfalse\endcsname 
 
\title{\bf A Study of the S=1/2 Alternating  
Chain using Multiprecision Methods} 
 
\author{T.Barnes$^1$, J.Riera$^2$ and D.A.Tennant$^3$} 
 
\address{ 
$^1$Physics Division, 
Oak Ridge National Laboratory,  
Oak Ridge, TN 37831-6373,  \\   
Department of Physics and Astronomy, 
University of Tennessee,  
Knoxville, TN 37996-1501, U.S.A. \\ 
$^2$Instituto de Fisica Rosario,  
2000 Rosario,  Argentina \\ 
$^3$Solid State Division, 
Oak Ridge National Laboratory,  
Oak Ridge, TN 37831-6393, U.S.A.  \\   
} 
 
\maketitle 
 
\begin{abstract} 
In this paper we present results  
for the ground state and low-lying excitations of the  
$S=1/2$ alternating Heisenberg 
antiferromagnetic chain. Our more conventional techniques include 
perturbation theory about the dimer limit and  
numerical diagonalization of  
systems of up to 28 spins. 
A novel application of multiple precision numerical diagonalization 
allows us to determine analytical perturbation series to high order; 
the results found using  
this approach include ninth-order perturbation series for the  
ground state energy and one magnon gap, which were previously known 
only to third order. We also give 
the fifth-order dispersion relation 
and third-order exclusive neutron scattering structure factor 
for one-magnon modes and 
numerical and analytical 
binding energies 
of $S=0$ and $S=1$ two-magnon bound states.  
  
\end{abstract} 
 
] 
\vspace{0.3cm}

\section{introduction}

The alternating Heisenberg chain (AHC) is a simple quantum spin system
that can be used to model the magnetic behavior of
a wide range of materials;
Table I gives some representative examples of alternating chains.
This model 
is a straightforward 
generalization of the uniform Heisenberg antiferromagnetic
chain, which is the most widely studied quantum spin system.
The uniform $S=1/2$ chain
has a gapless excitation spectrum with a known dispersion relation 
and a rather complicated 
ground 
state which is characterized by strong quantum fluctuations,
making it highly unstable to perturbations.

The alternating chain generalizes the uniform chain by alternating the
spin-spin interaction between two values, $J_1$ and $J_2$.
Since the alternating chain Hamiltonian is rotationally invariant with 
respect to spin, the total spin is a good quantum number, 
and the (antiferromagnetic) ground state is a spin singlet.
The translational symmetry of the uniform 1D chain however 
is reduced by dimerization, and the resulting system has a gap to the
first excited state, which has $S=1$. 
This lowest excitation is part of a ``one-magnon'' triplet band.
The alternating chain
has a rather complicated spectrum of states at higher energies,
including multimagnon continua and bound states. 

The alternating chain is of theoretical interest as a simple 1D 
isotropic quantum spin system with a gap, which presumably is qualitatively
similar to other more complicated 
systems such as integer-spin chains and even-leg $S=1/2$ spin ladders.
The approach to the uniform chain limit is also of interest 
as an example of critical behavior.
Finally, the alternating chain is useful as a model for the application of 
new numerical techniques such as the multiprecision approach introduced here.

The alternating chain Hamiltonian is realized in nature in many materials
that
have two important but structurally inequivalent 
superexchange paths that are spatially linked, 
so that a series of spin-spin 
interactions of strength $J_1-J_2-J_1-J_2...$ results. 
Examples of materials of this type are 
\vopo and \cuno and various 
aromatic free-radical compounds~\cite{NSM}.

Alternating chains may also arise as a result of the spin-Peierls effect.
In the 1D Heisenberg antiferromagnet 
a spatial dimerization of the ion positions along the chain gives
alternating interaction strengths, which results in a lowering
of the magnetic ground state energy.
There is a corresponding increase in the 
lattice energy (the phonon contribution) which dominates
at large distortions.
In the combined magnetic-phonon system an equilibrium is reached at a
spatial dimerization that minimizes the
ground state energy. This spontaneous dimerization is known as
the spin-Peierls effect, and the resulting magnetic Hamiltonian 
is an
alternating Heisenberg chain. 
Examples of spin-Peierls alternating chains in nature are
\cugeo and
\navons.

Much of the
recent interest in alternating chains arises from
the observation of a spin-Peierls effect in
CuGeO$_3$\cite{CGO_Nscat}. 
Many experimental
studies 
suggest an alternating chain 
interaction in \cugeo
(see for example \cite{CGO_2D} and references 
cited therein), although
interactions beyond nearest-neighbor 
are also thought to be important.
The observation of a two-magnon continuum in CuGeO$_3$ with an 
onset close to $2\Delta$ \cite{CGO_2D}, 
where $\Delta$ is the magnon energy-gap at the 
zone-center, has motivated recent theoretical studies of the continuum 
and two-magnon bound states in the AHC. 
Added impetus has come from neutron scattering 
studies of 
\vopo
\cite{our_vopo,our_vopowder}, 
which show that  
this material is dominantly an alternating chain and
provide evidence 
of a possible 
$S=1$ two-magnon bound state.

In this paper we present a detailed study of the AHC, including
many new results for the ground state and low-lying excitations.
We begin by introducing the model (Sec.II) and
reviewing previous studies (Sec.III). 
Perturbation theory about the dimer limit, which we find to be 
particularly well suited to studying the AHC, is 
introduced in Sec.IV.A and applied to the ground state energy
$E_0$, excitation gap $E_{gap}$ and
one-magnon dispersion $\omega (k)$ in Sec.IV.B.
Sec.IV.C summarizes analytical predictions for the critical behavior
of the gap and ground state energy. Sec.V presents our numerical results
for energies; Sec.V.A gives Lanczos results, and 
Sec.V.B introduces a new numerical method for abstracting
analytical perturbation series from high precision numerical results.
In Sec.V.B we use this new approach to
give series 
(based on $L=20$ diagonalization) to
$O(\alpha^9)$ for $E_0$, $E_{gap}$, and the zone-boundary energy
$E_{ZB}$. Previously these
series were known only to $O(\alpha^3)$. We also used this
multiple precision method
to determine the series expansion of $\omega (k)$
to $O(\alpha^5)$. These high-order formulae are accurate over a wide range
of alternations and should prove
useful to experimentalists. The critical
behavior is studied in Sec.V.C, and we present relationships between
the derivatives of $E_0$ and $E_{gap}$ at the critical point 
as well as comparing our results to the proposed scaling behavior 
of these quantities. 
Two-magnon bound states exist in the 
AHC, which is a convenient model for the study of this type of
excitation. We discuss the binding mechanism and
give second-order formulas for binding energies in Sec.VI. 
Since neutron scattering can give detailed information on the 
excitations of alternating chains, we
derive general expressions for the
exclusive neutron scattering structure factor 
${\cal S}(k)$ to a specific excitation
(Sec.VII.A). We apply these results 
to the excitation of the one-magnon band in Sec.VII.B, and use
the 
multiprecision method to calculate this ${\cal S}(k)$ to $O(\alpha^3)$.
A short discussion of the rather complicated 
neutron excitation of the two-magnon 
bound state band is given in Sec.VII.C. 
In Sec.VIII we present an illustrative application of our
new formulae to a real material, 
the alternating chain compound \vopons, for which
single-crystal neutron scattering data is available.
Finally we summmarize 
our results and present our conclusions in Sec.IX.

\section{the model} 
The AHC Hamiltonian is 
\begin{equation} 
H =    
\sum_{i=1}^{L/2} \  
J_1 \; 
{\vec S_{2i-1}} \cdot {\vec S_{2i}}   
+ J_2 \; {\vec S_{2i}} \cdot {\vec S_{2i+1}} \ .  
\end{equation} 
In this paper we impose periodic boundary conditions, 
with spins $1$ and $L+1$ identified. We usually
assume that   
$J_1 > J_2  >    0$, so we are in a regime 
of coupled antiferromagnetic dimers. 
 
We can also write this in terms of  
$J_1\equiv J$  
and the alternation  
$\alpha$, where $J_2 \equiv \alpha J$; 
\begin{equation} 
H =    
 \sum_{i=1}^{N_d=L/2}  
\ J \; {\vec S_{2i-1}} \cdot {\vec S_{2i}}   
+ \alpha J \;  {\vec S_{2i}} \cdot {\vec S_{2i+1}} \ .  
\label{Ham}
\end{equation} 
$N_d$ is the number of independent 
dimers or unit cells, which are coupled by the  
interaction $\alpha J$. 
 
An equivalent form often used in the discussion of 
spin-Peierls transitions writes this as interactions  
of strength ${\cal J}(1+\delta)$ and ${\cal J}(1-\delta)$, which are related 
to our definitions by 
${\cal J} =  (1+\alpha ) J/2$ 
and 
$\delta = (1-\alpha)/(1+\alpha)$ . 
 
For $\alpha =1$ this system is an isotropic, uniform,  
$S=1/2$  Heisenberg chain which has gapless excitations, 
and for $\alpha =0$ it reduces to uncoupled dimers with 
$E_{gap}=J$. 
Since this is an isotropic Hamiltonian with antiferromagnetic couplings, 
for $\alpha >0$  
we expect an $S=0$ (singlet) ground state and an $S=1$ (triplet) band 
of magnons as the first excitation. 
 
The geometry of our alternating chain is shown in Fig.1. 
Note that the unit cell has length $b$; this leads 
to a different set of momenta than the more familiar  
uniform chain, which has 
a unit cell of $a=b/2$. 
Since the Hamiltonian is invariant under translations by multiples of 
$b$,  
the allowed momenta are 
\begin{equation} 
k_n =  
{n \over L/4}  \cdot  {\pi \over b}    
\ . 
\label{kvalues}  
\end{equation} 
For $L/2$ even  
the index 
$n$ takes the values  
$n =0$, $\pm 1$, $\pm 2$, $\dots$, $\pm (L/4-1)$, $L/4$;  
for $L/2$ odd the series stops with $\pm int(L/4-1)$. 
There are $L/2=N_d$  
independent  
momenta  
because there 
are $N_d$ invariant translations of $H$. 
Positive and negative $k$ levels  
are degenerate as usual 
due to reflection symmetry.

\section{previous studies} 
 
Early  
numerical studies of the zero temperature alternating chain  
by Duffy and Barr \cite{DB} and Bonner and Bl\"ote 
\cite{BB} considered the ground state energy and triplet gap  
on chains of up to 10 and 12 spins 
respectively. They concluded that this system 
probably had a gap for any nonzero alternation.  
Duffy and Barr also gave results for the ground-state nearest-neighbor 
correlation function, magnetization in  
an external field, and triplet dispersion 
relation $\omega(k)$. 
A principal concern of Bonner and Bl\"ote and subsequent numerical work 
was to test the critical behavior of the uniform chain limit; 
analytical studies had 
predicted that the gap $E_{gap}/{\cal J}$ should open as $\delta^{2/3}$  
times logarithmic corrections for small 
alternation \cite{N}, and that  
the bulk-limit ground-state energy per spin expressed in terms of  
$\cal J$ and $\delta$, 
${\tilde e_0}=E_0/{\cal J}L = 2 e_0 / (1 + \alpha) $,  
should approach $1/4 - \ln(2)$ as  
$\delta^{4/3}$ times logarithmic corrections \cite{CF,BE}. 
The dependence of $\tilde e_0$ on $\delta$ is important in determining  
the existence of a 
spin-Peierls transition  
in an antiferromagnetic chain coupled to the phonon 
field~\cite{BB}. 
 
Numerical studies on larger systems  
were subsequently  
carried out by Soos {\it et al.} \cite{SKM}  
(to $L=26$ for $e_0$ 
and $L=21$ for $E_{gap}$) and  
Spronken {\it et al.} \cite{SFL} (to $L=18$).  
Spronken {\it et al.} supported the anticipated critical behavior. 
Soos {\it et al.} however considered much smaller $\delta$ and  
larger lattices and  
concluded that the expected asymptotic form was incorrect.  
This issue is unresolved and merits future study on much larger  
systems. 
 
More recent studies of the alternating chain model  
were motivated by experimental work on \cugeons \cite{CGO_Nscat,CGO_2D}. In particular 
the question of possible two-magnon bound states has been of interest; 
an analytical paper by 
Uhrig and Schulz \cite{US} anticipates an $S=0$ bound state 
for all $\delta$ and an $S=1$ bound state ``around $k=\pi/2$'' 
(our $k=\pi/b$, the zone boundary) ``for not too 
small $\delta$.''.  
Bouzerar {\it et al.} \cite{BKJ} similarly conclude that the $S=1$  
two-magnon bound state only exists for a range of $k$  
around the zone boundary. 
Fledderjohann and Gros \cite{FJ}  
have searched for evidence of such bound states 
in a numerical study of the 
structure factor $S(k,\omega)$ on chains of up to 
$L=24$, and conclude that an $S=1$ two-magnon bound state  
does indeed lie below 
the two-magnon continuum for all~$\delta$.

Numerical studies of the thermodynamic properties  
of the alternating chain  
have received much less attention.  
Duffy and Barr gave results for the internal energy, entropy, specific heat and 
magnetic susceptibility of an $L=10$ chain for a range of alternations. 
Diederix~{\it et al.}~\cite{DBGKP} specialized to the parameter 
$\alpha=0.27$   
appropriate for \cuno and gave results for the magnetization, susceptibility 
and entropy on systems of up to $L=12$. Barnes and Riera \cite{BR} gave 
results for the susceptibility on chains of up to $L=16$, and 
extrapolated to the bulk limit for values of $\alpha\approx 0.6$-$0.8$ 
considered appropriate for \vopons.

\section{Analytical results} 
 
\subsection{Dimer Perturbation Theory} 
 
Analytical results for the alternating chain 
can be derived using perturbation theory about the isolated dimer 
limit. For this purpose we partition the Hamiltonian into a dimer  
$H_0$  
and an interdimer interaction $H_I$, 
\begin{equation} 
H_0 =    
 \sum_{i=1}^{N_d}  
\ J \; {\vec S_{2i-1}} \cdot {\vec S_{2i}}   
\ ,  
\label{H_0}
\end{equation} 
\begin{equation} 
H_I =    
 \sum_{i=1}^{N_d}  
\alpha J  \;  {\vec S_{2i}} \cdot {\vec S_{2i+1}} \ .  
\label{H_I}
\end{equation} 
The single-dimer eigenstates of $H_0$ are an $S=0$ ground state  
$|\circ \rangle = (|\!\!\!\uparrow\downarrow\rangle -  
|\!\!\!\downarrow \uparrow\rangle )/\sqrt{2}$ 
with $E_0^{(dimer)} = -3J/4$ and 
an $S=1$ triplet of dimer excitations ``excitons'' 
$\{ |(+)\rangle, |(0)\rangle, |(-)\rangle \}$, with $E_1^{(dimer)} = + J/4$. 
We label these excitations by the dimer $S_z$, for example 
$|(0)\rangle =  
(|\!\uparrow\downarrow\rangle + |\!\downarrow \uparrow\rangle )/\sqrt{2}$. 
 
The ground state of the full $H_0$ is a direct product of $S=0$  
dimer ground states, 
\begin{equation} 
|\psi_0^{(0)}\rangle  =  
\prod_{m=1}^{N_d} |\; \circ_{m}\rangle   
\equiv |0\rangle, 
\label{H0gs}
\end{equation}  
with an energy of $E_0 = N_d \cdot E_0^{(dimer)} = - 3 J L / 8$. 
  
Similarly,  
the unperturbed one-magnon state with momentum $k$ and $S_z=+1$  
is given by 
\begin{equation} 
|\psi_1^{(0)}(k)\rangle^{(+)}  = {1\over \sqrt{N_d}} \sum_{m=1}^{N_d}  
e^{ikx_m}\; |(+)_m\rangle \ . 
\label{H0psi1}
\end{equation}  
(We will suppress the redundant polarization superscript on
$|\psi_1\rangle$ subsequently.)
We take the location $x_m$ of dimer $m$ to be the midpoint of 
the two spins. 
In this and similar state vectors, if the state of any dimer  
$n$ is not specified 
explicitly it is in the ground state $|\; \circ_n\rangle $. 
 
It is useful to derive the effect of $H_I$ on dimer product basis states. 
For example, operating with $H_I$  
on the $H_0$ ground state Eq.(\ref{H0gs}) gives 
\begin{equation} 
H_I \; |0 \rangle = 
{J\alpha \over 4}\sqrt{3} \sum_{m=1}^{N_d}  
\underbrace{ 
|(0,0)_{m,m+1}\rangle  
}_{\rm double\ excite} 
\end{equation} 
where  
\begin{displaymath} 
|(0,0)_{m,m+1}\rangle = \hskip 3cm  
\end{displaymath} 
\begin{equation} 
{1 \over \sqrt{3}}  
\bigg(  
  |(+)_m(-)_{m+1}\rangle   
- |(0)_m(0)_{m+1}\rangle   
+ |(-)_m(+)_{m+1}\rangle   
\bigg)  
\end{equation} 
is a state of two neighboring $S=1$ dimer excitons at dimer sites $m,m+1$ 
coupled to give $(S,S_z) = (0,0)$. 
Similarly the effect of  
$H_I$  
on a single $S_z=+1$  
exciton at dimer site $m$ gives 
\begin{displaymath} 
H_I \; |(+)_m \rangle = 
{\alpha J\over 4}  
\bigg\{  
\underbrace{ 
-|(+)_{m-1}\rangle 
-|(+)_{m+1}\rangle  
}_{\rm hop} 
\end{displaymath} 
\begin{displaymath} 
\underbrace{ 
- |(+)_{m-1} (0)_m\rangle + |(0)_{m-1} (+)_m\rangle 
}_{\rm excite} 
\end{displaymath} 
\begin{displaymath} 
\underbrace{ 
- |(+)_m (0)_{m+1}\rangle + |(0)_m (+)_{m+1}\rangle 
}_{\rm excite} 
\end{displaymath} 
\begin{equation} 
+ \sqrt{3}  
\sum_{m'=1}^{N_d} \!\! '  
\underbrace{ 
 |(+)_m (0,0)_{m',m'+1} \rangle  
}_{\rm double\ excite} 
\bigg\} \ . 
\end{equation} 
The prime on the sum indicates that  
all dimer sites  
represented in the state are distinct, 
so in this case $m'\neq m,m-1$. 
Evidently $H_I$ both translates the exciton (leading to momentum 
eigenstates) and couples it to two-exciton and three-exciton  
states of higher unperturbed 
energy. 
The specific polarization state 
$(|(+)(0)\rangle - |(0)(+)\rangle)/\sqrt{2}$  
is forced  
because this is the unique $|S=1,S_z=1\rangle$ combination of two $S=1$ dimers. 
We abbreviate this state as $|(1,1)_{m,n}\rangle$, specifying the  
$S_{total}$ 
and $S_{z\ total}$ and the excited dimers $m$ and $n$ ($m<n$),  
which gives 
the simplified form 
\begin{displaymath} 
H_I \; |(+)_m \rangle = 
- {\alpha J\over 4}  
\bigg\{ 
|(+)_{m-1}\rangle 
+|(+)_{m+1}\rangle 
\end{displaymath} 
\begin{displaymath} 
+ 
\sqrt{2} 
\bigg(  
 |(1,1)_{m-1,m}\rangle 
+ |(1,1)_{m,m+1}\rangle  
\bigg)  
\end{displaymath} 
\begin{equation} 
- \sqrt{3} \; 
\sum_{m'=1}^{N_d} \!\!'  
 |(+)_m (0,0)_{m',m'+1} \rangle  
\bigg\} \ . 
\end{equation} 

We can use this formalism 
to generate an expansion in  
$\alpha$ for the ground state 
and excitations and their matrix elements
using standard quantum mechanical
perturbation theory. These results are presented in the next section.
 
\subsection{Perturbative results for $E_0$, $E_{gap}$ and $\omega(k)$} 
 
The perturbative 
generalization of the ground state Eq.(\ref{H0gs})  
to $O(\alpha^2)$ is 
\begin{displaymath} 
|\psi_0\rangle = \eta_0 \Bigg[ 
|0\rangle  
+ \alpha  
\bigg\{ 
-{\sqrt{3}\over 8}  
\sum_{m=1}^{N_d} | (0,0)_{m,m+1}\rangle  
\bigg\}  
\end{displaymath} 
\begin{displaymath} 
+ \alpha^2  
\bigg\{ 
-{\sqrt{3}\over 32}  
\sum_{m=1}^{N_d} | (0,0)_{m,m+1}\rangle 
-{\sqrt{3}\over 32}  
\sum_{m=1}^{N_d} | (0,0)_{m,m+2}\rangle 
\end{displaymath} 
\begin{displaymath} 
+{1\over 16} \sqrt{2\over 3} \; 
\sum_{m=1}^{N_d} | (0,0)_{m,m+1,m+2}\rangle 
\end{displaymath} 
\begin{equation} 
+{3\over 128}  
\sum_{m,m'=1}^{N_d} \!\!\!\! '\; | (0,0)_{m,m+1}(0,0)_{m',m'+1}\rangle 
\bigg\}  
\Bigg]  \ , 
\label{psi0_2nd}
\end{equation} 
where $\eta_0=1-(3/128)\alpha^2 N_d$ is the  
$O(\alpha^2)$ 
normalization.  
Note that  
three- and four-exciton states appear at $O(\alpha^2)$.  
The four-exciton states encountered here are 
two  
$S=0$,  
$(0,0)_{m,m+1}$  
two-exciton pairs, again with the restriction on the sum $\sum '$ that 
no excitons 
overlap.  
The three-exciton 
state with $(S,S_z) = (0,0)$,  
\newpage
\begin{displaymath} 
|(0,0)_{m_1,m_2,m_3}\rangle = \hskip3cm 
\end{displaymath} 
\begin{displaymath} 
{1\over \sqrt{6}} 
\bigg( 
|(+)_{m_1}(0)_{m_2}(-)_{m_3}\rangle   
+|(0)_{m_1}(-)_{m_2}(+)_{m_3}\rangle   
\end{displaymath} 
\begin{displaymath} 
+|(-)_{m_1}(+)_{m_2}(0)_{m_3}\rangle   
-|(+)_{m_1}(-)_{m_2}(0)_{m_3}\rangle   
\end{displaymath} 
\begin{equation} 
-|(-)_{m_1}(0)_{m_2}(+)_{m_3}\rangle   
-|(0)_{m_1}(+)_{m_2}(-)_{m_3}\rangle   
\bigg) \ , 
\end{equation} 
is the unique $S=0$ combination of three 
spin-one objects at adjacent sites. 
 
Since the $O(\alpha^p)$ state determines the $O(\alpha^{2p+1})$ 
energy, we can in principle use Eq.({\ref{psi0_2nd}) 
to derive the ground state energy 
to $O(\alpha^5)$. This proves to be  
a rather intricate calculation. We have carried out this derivation 
of 
$e_0 \equiv  E_0/LJ$ analytically to 
$O(\alpha^4)$, with the result  
\begin{equation} 
e_0(\alpha ) = -{3\over 2^3 } - {3\over 2^6 } \, \alpha^2  
-{3\over 2^8} \, \alpha^3  
-{13\over 2^{12}} \, \alpha^4  
\ . 
\label{E0_dpt} 
\end{equation} 
This series 
was previously evaluated   
to $O(\alpha^3)$  
by Brooks Harris \cite{BH}. 
 
A similar $O(\alpha)$ 
generalization of the unperturbed  
$S~=~1$ one-magnon excitation Eq.({\ref{H0psi1}) gives  
\begin{displaymath} 
|\psi_1(k)\rangle  =  
\eta_1\, {1\over \sqrt{N_d}} 
\Bigg[ 
\sum_{m=1}^{N_d}  
e^{ikx_m}\;  
|(+)_m\rangle  
\end{displaymath} 
\begin{displaymath} 
+ \alpha  
\bigg\{ 
-{1\over 2\sqrt{2}}  
\sum_{m=1}^{N_d}  
(e^{ikb} + 1)\;  
e^{ikx_m}\;  
| (1,1)_{m,m+1}\rangle  
\end{displaymath} 
\begin{equation} 
-{\sqrt{3}\over 8} 
\sum_{m,m'=1}^{N_d} \!\!\!\! ' \;  
e^{ikx_m}\;  
| (+)_m (0,0)_{m'}\rangle  
\bigg\}   \Bigg] 
\label{psi1_1st} 
\ . 
\end{equation} 
The normalization $\eta_1=1$ to this order in $\alpha$.  
Taking the expected value of $H$ with this state  
gives a one-magnon dispersion relation  
of 
 
\begin{displaymath} 
{\omega(k) \over J} 
= 
\Big( 
1 
- {1\over 16}\, \alpha^2  
+ {3\over 64}\, \alpha^3  
\Big) 
- 
\Big( 
 {1\over 2}\, \alpha 
+ {1\over 4}\,\alpha^2  
- {1\over 32}\,\alpha^3  
\Big) 
\, \cos(kb)  
\end{displaymath} 
\begin{equation} 
- 
\Big( 
 {1\over 16}\,\alpha^2  
+ {1\over 32}\,\alpha^3  
\Big) 
\, \cos(2kb)  
- 
 {1\over 64}\,\alpha^3  
\, \cos(3kb)  \ . 
\label{Ek_dpt} 
\end{equation} 
The $O(\alpha^3)$ gap is therefore 
\begin{equation} 
{E_{gap}\over J} 
= 
 1  
- {1\over 2}\, \alpha   
- {3\over 8}\, \alpha^2 
+ {1\over 32}\, \alpha^3 
 \ . 
\label{Egap_dpt} 
\end{equation} 
These one-magnon energies were found previously to this order 
by Brooks Harris \cite{BH},  
and serve as a check of our $O(\alpha)$ one-magnon 
state Eq.({\ref{psi1_1st}).

\subsection{Critical behavior} 
 
As we approach the uniform chain the ground state energy and the gap  
are both expected to approach their limiting values as powers of 
$\delta$ times logarithmic corrections \cite{CF,BE}. The behavior near 
the critical point is usually discussed in terms of the variable 
 
\begin{equation} 
\delta = (1-\alpha)/(1+\alpha) \ , 
\end{equation} 
with  
\begin{equation} 
{\cal  J} = (1+\alpha)\, J  / 2 
\end{equation} 
fixed, 
so the alternating couplings are  
${\cal J} (1+\delta)$ and 
${\cal J} (1-\delta)$. 
These variables are more appropriate for a spin-Peierls 
system because a displacement of an intermediate ion by $O(\delta)$ should 
increase and decrease alternate couplings by approximately the same
amount. 
The ground state energy per spin 
relative to fixed ${\cal J}$ is 
$\tilde e_0(\delta) = 2 e_0 / (1+\alpha)$.

The critical behavior of the 
ground state
energy 
and 
singlet-triplet gap 
has been discussed 
by Cross and Fisher \cite{CF} and by Black and Emery \cite{BE}. The approach
used was to consider the properies of the Heisenberg chain
within a Luttinger-Tomonaga approximation, which involves 
a Jordan-Wigner transformation to a fermion representation of the
spin operators, and then replacing the cosinusoidal
fermion dispersion by a linear 
dispersion at the Fermi wavevector. This linear 
approximation is required to simplify the 
commutation relations between the density operators, allowing
the interacting fermion problem to be solved.
Renormalization techniques are then used to calculate the 
asymptotic behavior of various physical quantities within
this approximation. The approach makes uncontrolled 
approximations by neglecting states far from the 
Fermi surface and ignoring energy renormalization effects, so
it is unclear whether this
gives the correct critical behavior. 
The predicted asymptotic $\delta$ dependence is

\begin{equation} 
\lim_{\delta\to 0} \tilde e_0(\delta) - \tilde e_0(\delta=0)  \propto  
{\delta^{4/3} \over |\ln \delta \, |  }  
\end{equation} 
and
\begin{equation} 
\lim_{\delta\to 0} { E_{gap} \over {\cal J} }  \propto  
{\delta^{2/3} \over |\ln \delta \, |^{1/2}  } \ . 
\end{equation} 
We will compare these predictions with our numerical results 
in the next section.

One may derive some relations between energies and their derivatives
near the critical point 
from a simple identity satisfied by the alternating chain Hamiltonian 
Eq.(\ref{Ham}). 
Note the proportionality relation 
\begin{equation} 
H(J,\alpha J) = \alpha \cdot H(\alpha^{-1} J,J) \ , 
\end{equation} 
which implies for any energy eigenvalue 
\begin{equation} 
{E_n(\alpha)\over J}  = \alpha \cdot {E_n(\alpha^{-1})\over J}  \ . 
\label{recip} 
\end{equation} 
 
Assuming that there are no singularities on the real axis except at  
$\alpha=1$, we may differentiate this relation with respect to 
$\alpha$ elsewhere.  
As it is expected that the singularity in 
$e_0(\alpha) = E_0/JL$  
at the critical point $\alpha~=~1$ is higher order than linear,  
$d e_0(\alpha) / d \alpha$ should be  
well defined everywhere;  differentiating 
Eq.({\ref{recip}) 
with $n=0$ therefore leads to 
\begin{equation} 
{ d e_0(\alpha) \over d \alpha}\bigg|_{\alpha=1} = {1\over 2}\, 
e_0(\alpha=1) = {1\over 8} - {\ln(2)\over 2}  = -0.22157... \ . 
\end{equation} 
 
This is consistent with the expectation that the scaled  
$\tilde e_0(\delta)$ has zero slope in $\delta$ as we approach the 
critical point. To see this, note that $\tilde e_0(\delta) =  
2 e_0(\alpha)/(1+\alpha)$, 
so 
\begin{equation} 
{ d \tilde e_0(\delta) \over d \delta} =  
e_0(\alpha) - {2\over (1+\delta)}\,   
{ d e_0(\alpha) \over d \alpha}  
\ , 
\end{equation} 
and as we approach the critical point 
\begin{equation} 
\lim_{\delta\to 0} { d \tilde e_0(\delta) \over d \delta} =  
\lim_{\alpha\to 1} \ \ e_0(\alpha) - 2 \, 
{ d e_0(\alpha) \over d \alpha}  
 = 0 \ . 
\end{equation} 
Successive derivatives of Eq.(\ref{recip}) can be used to infer relations 
between higher derivatives of $e_0(\alpha)$ (or other energy eigenvalues) 
as one approaches  
$\alpha=1$. 
 
\section{Numerical Methods} 
 
\subsection{Lanczos results}
 
Direct numerical diagonalization  
of moderately large systems is possible for the 
$S=1/2$   
alternating 
chain.  
Here we used a Lanczos method
\cite{josesref} 
to obtain ground state and one-magnon energies on 
$L=4n$ lattices up to $L=28$. 
Motivated by previous numerical 
studies, we extrapolate these energies to bulk limits 
using a simple exponential-and-power estimate for the 
finite size dependence, 
\begin{equation} 
f(\alpha,L) = f(\alpha) + c_1\; {\exp(-L/c_2) \over L^p} 
\label{lanc_extrap} 
\end{equation} 
where $p=1$ for energy gaps $(f=E_n-E_0)$ and $p=2$ for the ground state 
energy per spin $(f=E_0/L)$. 
We determined the finite-lattice energies to 
about 14 place accuracy, and fitted  
the  
$L=4(n-2), 4(n-1)$ and $4n$ results to these asymptotic forms. 

The resulting bulk-limit ground state energy is
shown in Fig.2 and
presented  
in Table II, to nine figure 
accuracy for the smaller $\alpha$ values. 
For larger $\alpha$ 
we include the change in $e_0(\alpha)$ between
$L=16,20,24$ and $L=20,24,28$  
extrapolations 
in parenthesis after the tabulated $L=20,24,28$ result, as an error
estimate.
These numerical energies 
provide an accurate check of the perturbative formulas 
Eq.(\ref{E0_dpt}) 
and 
Eq.(\ref{E0_mp}).

We also used Lanczos diagonalization on lattices 
up to $L=28$ to determine the singlet-triplet gap and zone boundary
energy. 
These results are given in Table II and shown in Fig.3,  
again with a systematic  
error estimate that is 
the discrepancy between $L\leq 24$ and $L\leq 28$ extrapolations.  
(The + sign indicates 
that the bulk-limit gap estimate increased with increasing $L$.)  
Our Lanczos results are again consistent with the perturbative 
expansions 
Eq.(\ref{Ek_dpt})  and
Eq.(\ref{Egap_dpt}).
The higher-order multiple precision series
Eq.(\ref{Egap_mp}) and
Eq.(\ref{EZB_mp}) are as expected found to be in agreement to much
higher accuracy.

\subsection{High-order series from multiple precision} 
 
One may use numerical diagonalization combined with  
multiple precision programming to determine analytical  
perturbation series to high order.  
For this novel application of numerical methods to spin systems 
we employed the multiprecision package MPFUN developed by D.Bailey  
\cite{MPFUN}, applied to our Fortran code 
for low-lying eigenvectors of  
the alternating chain using the ``modified Lanczos''  
method \cite{mLmethod}.  
We typically generated energies to 300 significant figures 
with $\alpha=10^{-30}$ (and to 400 figures for $L=20$), which  
allowed the perturbation expansion coefficients to be read directly 
from the numerical energies  
as rational fractions. This was possible in part because the energy  
denominators involved simple powers of small integers that could be 
anticipated. 
The limiting order in this approach is determined 
by the size of the  
system one can diagonalize, since the periodic boundary conditions 
introduce  
$O(\alpha^{L/2})$ 
finite-lattice corrections to energies. 
This gave a limit of $O(\alpha^9)$ for the order of the  
bulk limit expansion 
that could be determined from the  
largest system we diagonalized with multiple precision, $L=20$. 
 
The multiprecision package was 
implemented on  
a Pentium PC, a DEC Alpha and a Sun 450.  
Execution times for $E_0$ 
at this level of precision  
were approximately 6~CPU hours 
for $L=16$ 
on the DEC Alpha 
and 100 CPU hours for $L=20$ on the  
Sun 450. 
 
The $O(\alpha^9)$  
series for the ground state energy per spin determined 
in this manner is 
 
\begin{displaymath} 
e_0(\alpha ) = -{3\over 2^3 } - {3\over 2^6 } \alpha^2  
-{3\over 2^8} \alpha^3  
-{13\over 2^{12}} \alpha^4  
-{89\over 2^{14}\cdot 3} \alpha^5  
\end{displaymath} 
\begin{displaymath} 
-{463\over 2^{17}\cdot 3} \alpha^6  
-{7\cdot 61 \cdot 191 \over 2^{22}\cdot 3^3} \alpha^7  
-{11\cdot 139\cdot 271 \over 2^{21}\cdot 3^4 \cdot 5} \alpha^8  
\  
\end{displaymath} 
\begin{equation} 
-{107 \cdot 22005559 \over 2^{30}\cdot 3^5 \cdot 5^2} \alpha^9  
\label{E0_mp}
\end{equation} 
and the  
$O(\alpha^9)$  
series for the gap to the $k=0$ one-magnon state is  
\begin{displaymath} 
{E_{gap}\over J} 
=  
 1  
- {1\over 2}\, \alpha   
- {3\over 2^3 }\, \alpha^2 
+ {1\over 2^5}\, \alpha^3 
- {5\over 2^7 \cdot 3}\, \alpha^4   
\end{displaymath} 
\begin{displaymath} 
- {761 \over 2^{12} \cdot 3 }\, \alpha^5 
+ {(11)^2 \cdot 157 \over 2^{16} \cdot 3^3}\, \alpha^6   
+ {21739 \over 2^{18} \cdot 3^3 }\, \alpha^7  
\end{displaymath} 
\begin{equation} 
- {107  \cdot 283  \cdot 7079 \over 2^{24} \cdot 3^4 \cdot 5}\, \alpha^8  
+ {1307  \cdot 9151183  \over 2^{28} \cdot 3^6 \cdot 5^2}\, \alpha^9  \ . 
\label{Egap_mp}
\end{equation} 
The  
zone-boundary $(k=\pi / b)$ energy of the one-magnon state 
relative to $E_0$ 
to this order is 
\begin{displaymath} 
{E_{ZB}\over J} 
=  
 1  
+ {1\over 2}\, \alpha   
+ {1\over 2^3 }\, \alpha^2 
- {1\over 2^5 \cdot 3 }\, \alpha^4 
- {83\over 2^{12} \cdot 3}\, \alpha^5 
\end{displaymath} 
\begin{displaymath} 
- {71 \cdot 149 \over 2^{16} \cdot 3^3 }\, \alpha^6 
- {6373\over 2^{14} \cdot 3^4 }\, \alpha^7 
- {19 \cdot 128461\over 2^{24} \cdot 3^2\cdot 5 }\, \alpha^8  
\end{displaymath} 
\begin{equation} 
- {41 \cdot 256687901 \over 2^{28} \cdot 3^6\cdot 5^2 }\, \alpha^9 \ .  
\label{EZB_mp}
\end{equation} 
 
We can also use multiprecision methods to determine the  
one-magnon dispersion relation $\omega(k)$, parametrized by   
\begin{equation} 
{\omega(k)\over J}  =   
\sum_{\m=0}^{\infty} \, a_\m(\alpha) \, \cos(\m kb) \ . 
\label{def_Fourier} 
\end{equation} 
On a finite lattice, momenta are only defined at  
the $N_d=L/2$ independent values of 
Eq.({\ref{kvalues}). (See also Fig.1b.) 
The relation 
$\omega(k) = \omega(-k)$ further reduces this to a total of 
{\it int}$(N_d/2) + 1 $ 
independent  
lattice energies. 
These  
can  
be expanded as power series in $\alpha$, and we 
again encounter 
finite 
lattice artifacts in these expansions beginning at 
$O(\alpha^{N_d})$.  
 
The use of the lattice 
$\{ \omega(k_n) \} $ 
to determine the Fourier coefficients 
in 
Eq.({\ref{def_Fourier}) 
is nontrivial  
because there are infinitely many coefficients 
but only 
{\it int}$(N_d/2) + 1 $ 
lattice energies. 
We can proceed by determining the  
{\it int}$(N_d/2) + 1 $ 
{\it finite lattice} Fourier coefficients 
$\{ \hat a_\m (\alpha,L ) \} $,  
$\m =0,1,...,${\it int}$(N_d/2)$, defined by 
\begin{equation} 
{\omega(k_n) \over J}\big|_{L}    
=  \sum_{\m =0}^{{\it int}(N_d/2)} \, \hat a_\m(\alpha,L) \,  
\cos(\m k_nb)   
\label{def_latt_Fourier} 
\end{equation} 
where $k_n b = 2\pi n / N_d$.

We can invert this using a general result for the sum 
of a product of cosines over the discrete lattice momenta. 
We shall now assume  
$N_d$ is even, so there are singly degenerate  
$k=0$ and $k=\pi/b$ points in addition to the doubly 
degenerate values $k=\pm 2\pi / N_d b, \pm 4\pi / N_d b, ...$. 
We can translate the $k=0$ and negative $k$ values by  
$2\pi / b$, so a sum over lattice $k$ values becomes a sum over 
$n=1,2,...,N_d$, with $k_n = 2\pi n / N_d b$.  
The summed product of cosines is 
\begin{displaymath} 
{1\over N_d}  
 \sum_{\m=1}^{N_d} \,   
\cos(2\pi \m n / N_d ) \   
\cos(2\pi \m n' / N_d )  =   
\hskip 1cm 
\end{displaymath} 
\begin{equation} 
{1\over 2} \Big( 
\delta_{{\rm mod}(n-n',N_d), 0}  +  
\delta_{{\rm mod}(n+n',N_d), 0}   
\Big)   
\ . 
\label{cosprod} 
\end{equation} 
On multiplying 
Eq.(\ref{def_latt_Fourier}) 
by $\cos(2\pi n \m' / N_d)$ and summing over 
$n$, we therefore find the lattice Fourier coefficients 
\begin{displaymath} 
\hat a_\m(\alpha, L) =  
\hskip 5cm 
\end{displaymath} 
\begin{equation} 
{2\over N_d} {1\over  
( 1  
+ \delta_{\m,0}  
+ \delta_{\m,N_d/2}  
) } 
\sum_{n=1}^{N_d}  
{\omega(k_n) \over J}\big|_{L} \cdot  \cos(2\pi \m n / N_d ) \ . 
\label{latt_coeffs} 
\end{equation} 
 
If we assume more generally that the $\{ \omega (k_n) \} $ are 
samplings of the continuous function given by Eq.(\ref{def_Fourier}), 
and invert the Fourier expansion using Eq.(\ref{cosprod}) we find 
\begin{displaymath} 
{2\over N_d}   
\sum_{n=1}^{N_d}  
{\omega(k_n) \over J}\big|_{L} \cdot  \cos(2\pi n \m' / N_d ) = 
\hskip 1cm 
\end{displaymath} 
\begin{equation} 
\sum_{\m=0}^{\infty} a_\m  
\; \Big( 
\delta_{{\rm mod}(\m-\m',N_d), 0}  +  
\delta_{{\rm mod}(\m+\m',N_d), 0}   
\Big)  \ . 
\label{bulk_coeffs} 
\end{equation} 
As anticipated this constitutes 
{\it int}$(N_d/2) + 1 $ constraints (the number of independent choices for 
$\m'$) on the infinite set of Fourier coefficients. 
 
As an illustration, for $\m'=0$ 
Eqs.(\ref{latt_coeffs}) and
(\ref{bulk_coeffs}) 
imply 
\begin{displaymath} 
\hat a_0 = {1\over N_d}   
\sum_{n=1}^{N_d}  
{\omega(k_n) \over J}\big|_{L}  = 
\hskip 3cm 
\end{displaymath} 
\begin{displaymath} 
{1\over 2} \; \sum_{\m=0}^{\infty} a_\m  
\Big( 
\delta_{{\rm mod}(\m,N_d), 0}  +  
\delta_{{\rm mod}(\m,N_d), 0}   
\Big) =  
\end{displaymath} 
\begin{equation} 
a_0 + a_{N_d} + a_{2N_d} + ... \ . 
\hskip 1cm 
\end{equation} 
Since we know that the perturbative series for 
both the lattice $\{ \hat a_\m \} $ and bulk-limit  
$\{  a_\m \} $ coefficients 
begin at  
$O(\alpha^\m)$, the expansions of 
$a_0$ and $\hat a_0$ must be identical
until we encounter 
contributions 
from higher-order coefficients; for $a_0$ these 
begin at $O(\alpha^{N_d})$ due to $a_{N_d}$.  
This is also the order at which finite-lattice artifacts appear 
in the energies, 
so there has been no loss of order reached in  
the $a_0$ expansion beyond the 
usual finite-lattice limitation. 
An equivalent conclusion follows for the Fourier coefficient 
$a_{N_d/2}$.  
 
For the other Fourier modes the order of perturbation 
theory to which the
bulk-limit coefficients 
are determined
is reduced by the contributions of higher-order  
coefficients. For $\m'=1$ for example we find 
 
\begin{displaymath} 
\hat a_1 = {2\over N_d}   
\sum_{n=1}^{N_d}  
{\omega(k_n) \over J}\big|_{L} \cdot  \cos(2\pi n / N_d ) = 
\hskip 2cm 
\end{displaymath} 
\begin{displaymath} 
\sum_{\m=0}^{\infty} a_\m  
\; \Big( 
\delta_{{\rm mod}(\m-1,N_d), 0}  +  
\delta_{{\rm mod}(\m+1,N_d), 0}   
\Big) =  
\hskip 0.5cm 
\end{displaymath} 
\begin{equation} 
a_1 +  
a_{N_d+1} +  
a_{2N_d+1} + ... +  
a_{N_d-1} +  
a_{2N_d-1} + ... \ . 
\end{equation} 
The contributions of the two delta functions that combined 
in the previous $\m'=0$ example are now distinct, 
so there is an $O(\alpha^{N_d-1})$ difference due to 
$a_{N_d-1}$ between the 
known lattice series for $\hat a_1$ and the bulk-limit $a_1$ series. 
The worst case is $m'=N_d/2 - 1$; 
the lattice and bulk-limit coefficients are then related by 
 
\begin{displaymath} 
\hat a_{N_d/2-1} =  
\hskip 4cm 
\end{displaymath} 
\begin{equation} 
a_{N_d/2-1} +   
a_{3N_d/2-1} + ... + 
a_{N_d/2+1} +  
a_{3N_d/2+1} + ...  
\ , 
\end{equation} 
so the series for the bulk-limit 
$a_{N_d/2-1}$, 
which begins at  
$O(\alpha^{N_d/2-1})$,  
cannot be determined beyond  
$O(\alpha^{N_d/2})$ due to the presence of $a_{N_d/2+1}$.  
Thus we conclude that the perturbative expansion 
of the bulk-limit dispersion relation Eq.(\ref{def_Fourier}) can only be  
{\it completely} 
determined   
to $O(\alpha^{N_d/2})$ from finite lattice data. The attainable order 
depends on the mode, and is in general 
$O(\alpha^{N_d-1-{\rm mod}(\m,N_d/2)})$ 
for $a_\m$ with $N_d/2 \geq \m \geq 0$.

It follows that  
the complete set of 
bulk-limit Fourier coefficients 
$\{ a_\m \} $ 
is 
uniquely determined 
just to  
$O(\alpha^5)$  
by the diagonalization of the $L=20$ lattice.  
These coefficients, which confirm 
and continue 
the one-magnon dispersion relation of Brooks Harris, 
Eq.(\ref{Ek_dpt}), are  
 
\begin{equation} 
\begin{array}{lrrrrrrr} 
a_0   
& =  
& 1  
&   
& - {1\over 16}\, \alpha^2  
& + {3\over 64}\, \alpha^3  
& + {23\over 1024}\, \alpha^4   
& - {3\over 256}\, \alpha^5   
\\ 
\\ 
a_1  
& =   
& 
& - {1\over 2}\, \alpha 
& - {1\over 4}\,\alpha^2  
& + {1\over 32}\,\alpha^3  
& + {5\over 256}\, \alpha^4 
& - {35\over 2048}\, \alpha^5   
\\ 
\\ 
a_2  
& =   
& 
& 
& - {1\over 16}\,\alpha^2  
& - {1\over 32}\,\alpha^3  
& - {15\over 512 }\, \alpha^4 
& - {283\over 18432}\, \alpha^5   
\\ 
\\ 
a_3  
& =   
& 
&  
&  
& - {1\over 64}\,\alpha^3  
& - {1\over 48}\, \alpha^4 
& - {9\over 1024}\, \alpha^5   
\\ 
\\ 
a_4  
& = 
&   
&   
&   
&   
& - {5\over 1024}\, \alpha^4 
& - {67\over 9216}\, \alpha^5   
\\ 
\\ 
a_5  
& = 
&   
&   
&   
&   
&  
& - {7\over 4096}\, \alpha^5   
\end{array} 
\ . 
\label{ans_Fourier} 
\end{equation} 
 
The fifth-order one-magnon dispersion relation given by 
Eqs.(\ref{def_Fourier},\ref{ans_Fourier}) 
is 
shown for a range of alternations in Fig.4.  
 
\subsection{Numerical results for critical behavior}
 
The expected behavior  
of $E_{gap}/{\cal J} $  
and  
$\tilde e_0 (\delta) = E_0/{\cal J} L  
= 2 e_0(\alpha)/(1+\alpha)$  
as we approach 
the critical point $\delta=0$ is \cite{CF,BE} 
\begin{equation} 
\lim_{\delta\to 0} \tilde e_0(\delta) - \tilde e_0(\delta=0)  \propto  
{\delta^{4/3} \over |\ln \delta \, |  } 
\label{E0_asymp} 
\end{equation} 
and 
\begin{equation} 
\lim_{\delta\to 0} {E_{gap}\over {\cal J}}    
\propto  
{\delta^{2/3} \over |\ln \delta \, |^{1/2}  } \ . 
\label{Egap_asymp} 
\end{equation} 
We will now compare these theoretical asymptotic forms with our numerical 
results for 
$E_{gap}/{\cal J}$ 
and 
$\tilde e_0(\delta)$ 
at small 
$\delta$.  
 
Taking the ground state energy per spin first, in Fig.5a we show 
a logarithmic plot of $\tilde e_0$ versus $\delta$;  
evidently the smaller $\delta$ values do support approximate 
power-law behavior. A fit of the three smallest $\delta$ 
points to the form $c_1\delta^p$ gives $c_1=0.3632$ and $p=1.412$, 
apparently
consistent with $\delta^{4/3}$ times logarithmic corrections. 
The anticipated theoretical form  
$c_1 \delta^{4/3} / |\ln \delta \, | $ however does not give an especially 
good fit, and is actually  
worse than a pure $4/3$ power law. 
A $4/3$ power fit to the three smallest $\delta$ points 
gives $c_1=0.3134$, which is reasonably accurate for larger $\delta$ as well;
this fit is shown in Fig.5a. 
 
The $\delta$ dependence we observe for the singlet-triplet gap
scaled by ${\cal J}$ 
is shown in Fig.5b.
The approximate linearity of the log-log plot suggests
power-law behavior over the full range of $\delta$ considered.
A fit of all points to
$c_1\delta^p$ gives $c_1=1.999$ and $p=0.7497$,
suggesting a power law of $\delta^{3/4}$ rather than 
the theoretical $\delta^{2/3}$ of Eq.(\ref{Egap_asymp}). 
Fitting $c_1 \delta^{3/4}$ to the three smallest $\delta$ points
gives $c_1=2.003$, which is shown in Fig.5b. Evidently this is
a remarkably good fit. The surprising accuracy of this form for
large alternation 
can be understood by noting that the similar 
function 
$E_{gap} / {\cal J} = 2 \, \delta^{3/4}$, 
corresponding to
$E_{gap} / J =  (1-\alpha)^{3/4}(1+\alpha)^{1/4}$, 
gives the correct $E_{gap}/J$ 
power series 
Eq.(\ref{Egap_dpt})
to $O(\alpha^2)$. 

The form  
$c_1 \delta^{2/3} / |\ln \delta \, |^{1/2} $ found by Black and Emery
gives a much less accurate description of our data. Thus we appear to
support a different asymptotic power law than expected for the singlet-triplet
gap, although we cannot continue to very small
$\delta$ to see if this discrepancy persists.
The behavior of the ground state energy and gap 
much closer to the critical point would be an interesting
topic for a detailed investigation on large lattices, using for example
the DMRG approach \cite{SteveW}.

\section{Two-Magnon Bound States} 
 
Recent theoretical work  
\cite{US,BKJ,FJ,PRHBH} 
motivated by neutron scattering studies of 
CuGeO$_3$ \cite{CGO_2D} 
has led to considerable interest in  
two-magnon bound states in dimerized quantum spin systems. 
The existence of such bound states in the alternating chain model 
appears very likely, since the 
interdimer interaction $H_I$, Eq.(\ref{H_I}),  
gives an attractive diagonal potential 
energy between two adjacent excited dimers if their total spin is $S=0$ or 
$S=1$. To see this, note the effect of $H_I$ on a state of adjacent  
$(+)(0)$ dimer 
excitons;  
\begin{displaymath} 
H_I  
|  
\circ 
(+) 
(0) 
\circ 
\rangle 
=  
{\alpha J\over 4}  
\bigg\{ 
\underbrace{ 
|  
\circ 
(0) 
(+) 
\circ 
\rangle 
}_{\rm transpose} 
\end{displaymath} 
\begin{displaymath} 
\underbrace{ 
- 
|  
(+) 
\circ 
(0) 
\circ 
\rangle 
- 
|  
\circ 
(+) 
\circ 
(0) 
\rangle 
}_{\rm hop} 
\end{displaymath} 
\begin{displaymath} 
\underbrace{ 
+ 
|  
\circ 
(+) 
\circ 
\circ 
\rangle 
+ 
|  
\circ 
\circ 
(+) 
\circ 
\rangle 
}_{\rm deexcite} 
\end{displaymath} 
\begin{displaymath} 
\underbrace{ 
+ 
|  
(+) 
(0) 
(0) 
\circ 
\rangle 
- 
|  
(0) 
(+) 
(0) 
\circ 
\rangle 
}_{\rm excite} 
\end{displaymath} 
\begin{displaymath} 
\underbrace{ 
+ 
|  
\circ 
(+) 
(+) 
(-) 
\rangle 
- 
|  
\circ 
(+) 
(-) 
(+) 
\rangle 
}_{\rm excite} 
\end{displaymath} 
\begin{equation} 
+ \sqrt{3} \sum_{m'=1}^{N_d} \!\! '  
\underbrace{ 
| (0,0)_{m',m'+1} \dots 
(+) 
(0) 
\dots  
\rangle 
}_{\rm double\ excite} 
\bigg\}  
\ . 
\end{equation} 
Again in $\sum'$ the index $m'$ takes on all values 
that do not superimpose a dimer excitation on the already excited 
$(+)$ or $(-)$ sites.
 
The analogous expressions for other polarization states may be  
combined to give the effect of $H_I$ on nearest neighbor 
dimer excitons with definite total spin. There is a diagonal interaction 
due to the ``transpose'' matrix element and others that retain two 
neighboring excitons. This static inter-exciton potential is 
\begin{equation} 
\langle (S,S_z)_{m,m+1}| 
H_I 
| 
(S,S_z)_{m,m+1} 
\rangle  
 =  
\cases{ 
+J\alpha/4 & $S=2$; \cr 
-J\alpha/4 & $S=1$; \cr 
-J\alpha/2 & $S=0$. \cr} 
\label{EBstatic}  
\end{equation} 
 
This suggests that at large alternation (small $\alpha$) we should find  
two-magnon bound states with $S=0$ and $S=1$, with binding energies  
of approximately $J\alpha/2$ and $J\alpha/4$ respectively.  
(This only applies to $k=\pi/b$; at  
other $k$ values there is a complication that modifies this result, 
as noted below.) 
 
We can again use numerical results with small coupling $\alpha$ 
to establish the higher-order perturbation series for properties of 
these bound states. Unfortunately this is a much more difficult  
numerical problem than the study of the  
ground state and one-magnon levels, 
so here we give only a few preliminary results. 
 
The binding energies $E_B$ of the $S=0$ and $S=1$ bound states are defined by 
\begin{equation} 
E_B(k) = \min_{k_1+k_2 = k}\Big(\omega(k_1) + \omega(k_2)\Big)  
- \Big( E(k) - E_0\Big) \ . 
\label{EBdefined}
\end{equation} 
(The $\omega$ sum  
gives the onset of the two-magnon 
continuum at $k$.) Assuming that the continuum  
onset at $k=\pi/b$ is given   
by $k_1=0$ and $k_2=\pi/b$, we 
find that  
the  
two-magnon binding energies  
to $O(\alpha^2)$ on $L\geq 12$ lattices 
are 
\begin{equation} 
{E_B(k=\pi/b) \over J} \; = \;  
\cases{ 
{1\over 4} \alpha - {13\over 32}\alpha^2 & $S=1$, \cr 
\cr 
{1\over 2}\alpha - {7\over 16} \alpha^2 & $S=0$. \cr} 
\label{EB_mp}
\end{equation} 
Our numerical result for the bulk-limit  
binding energies at $k=\pi/b$ (Table II and Fig.6)  
are clearly consistent with 
these perturbative results at small $\alpha$. 
It is interesting that the binding 
is much weaker near $k=0$, for example the expansion of the 
$S=0$ binding energy at $k=0$ appears to begin at 
$ O(\alpha^2)$. 
 
The determination of these perturbative binding energies analytically at 
general $k$ is a complicated problem because the  
two-exciton sector is a manifold of degenerate states 
under $H_0$. Determining the appropriate basis states within this 
degenerate manifold requires diagonalization of the ``hopping'' 
part of $H_I$. This is only straightforward at $k=\pi/b$, 
where this hopping amplitude vanishes; as an illustration, 
for adjacent $(+)(0)$ excitons we find 
\begin{equation} 
\langle\; (+)\circ (0) \circ \ ; k  | H_I |\circ (+)(0)\circ\ ; k \rangle 
\propto \cos(kb/2) 
\ .  
\end{equation} 
Thus $k=\pi/b$ bound states do not mix with the two-exciton continuum to 
leading order, and remain relatively localized.  
 
Except near $k=\pi/b$ we expect the coupling between 
nearest-neighbor and separated excitons to be very important, 
because the energy denominator is $O(\alpha J)$, which is the 
same order as the hopping matrix element. This implies  
that there are  
corrections to the bound-state wavefunction of order
$O(\langle H_I \rangle / \Delta E) =  
O(\alpha J)/O(\alpha J) = O(\alpha^0)$  
and 
corrections to the  
static binding energy 
Eq.(\ref{EBstatic}) 
of order
$O(|\langle H_I \rangle|^2 / \Delta E) =  
O(\alpha)$  
except at  
$k=\pi/b$.

At the extreme point $k=0$ we see no evidence 
for a $S=1$ bound state in our numerical extrapolation of
Lanczos data. Although we do see evidence of an
$S=0$ bound state at $k=0$, it is quite weakly bound 
and (in consequence) has a very 
extended spatial wavefunction.  
The difference between $k=0$ and $k=\pi/b$ bound state wavefunctions
was quite evident in the 
finite-size effects seen in our numerical extrapolation. 
 
In a preliminary numerical study of a truncated system of  
zero-, one- and two-exciton states on an $L=200$ lattice we 
find that the attractive potentials in 
Eq.(\ref{EBstatic}) 
are strong enough 
to form an $S=0$ bound state for all $k$, but apparently the $S=1$ 
bound state exists only for a range of $k$ around $\pi/b$. 
Similar conclusions have been reported by Uhrig and Schulz \cite{US}  
and Bouzerar {\it et al.} \cite{BKJ}.  
 
These results are especially relevant to  
the alternating-chain 
material \vopons,
``VOPO''. 
An 
excitation has been observed in VOPO 
just below the two-magnon continuum, 
which has been cited as a possible two-magnon bound state 
\cite{our_vopo,our_vopowder}. Since this peak is seen clearly 
for a range of $k$ including  
$k\approx  2\pi / b$ (equivalent to the $k=0$ point where we
find no $S=1$ bound state), it is inconsistent with 
the expectations of the alternating chain model for such a bound state. 
If mode observed in VOPO 
is indeed a two-magnon bound state, its persistence to
$k\approx 2\pi / b$  is 
presumably due to additional interactions.

\section{Neutron scattering structure factor} 
 
\subsection{General results}

Identification of the magnetic excitations predicted by the  
alternating chain model will be facilitated by  
estimates of their couplings to external probes, such as 
photons (especially for $S=0$ states, through Raman scattering) 
and neutrons (for $S=1$ states). Here we present perturbative and 
numerical results for the 
neutron scattering structure factor to the first two $S=1$ excitations 
in the alternating chain, which are  
the one-magnon mode and the $S=1$  
two-magnon bound state. 
 
The  
neutron scattering cross section is proportional to the structure factor, 
which in the Heisenberg picture is 
\begin{displaymath} 
{\rm S}_{mm'}(\vec k, \omega) =  
{1\over 2\pi} \int_{-\infty}^{\infty} dt  
\; e^{i \omega t} 
\; e^{i \vec k \cdot (\vec x_i - \vec x_j ) }  
\end{displaymath} 
\begin{equation} 
\sum_{sites\ i,j}  
\langle \psi_0 | S_m^{\dagger}(\vec x_j,t) S_{m'}(\vec x_i,0) |\psi_0\rangle 
\ . 
\end{equation}  
 
We can insert a complete set of eigenstates $\{|\psi_N\rangle\}$ of the 
full $H$ between the spins in this matrix element and write 
the spin structure factor as a sum over  
{\it exclusive structure factors},  
\begin{equation} 
{\rm S}_{mm'}(\vec k, \omega) =  
\sum_{N} {\rm S}_{mm'}^{\psi_0\to\psi_N}(\vec k, \omega)   
\end{equation}  
where 
\begin{displaymath} 
{\rm S}_{mm'}^{\psi_0\to\psi_N}(\vec k, \omega) =  
\sum_{sites\ i,j}  
{1\over 2\pi} \int_{-\infty}^{\infty} dt  
\; e^{i \omega t} 
\; e^{i \vec k \cdot (\vec x_i - \vec x_j ) }  
\end{displaymath} 
\begin{equation} 
\langle \psi_0 | S_m^{\dagger}(\vec x_j,t)  
|\psi_N\rangle\langle\psi_N|  
S_{m'}(\vec x_i,0) |\psi_0\rangle \ .  
\end{equation}  

Each exclusive
structure factor 
S$^{\psi_0\to\psi_N}_{mm'}(\vec k,\omega)$, 
gives
the intensity of scattering from 
$|\psi_0\rangle $ 
to a specific excited state 
$|\psi_N\rangle $, 

The Heisenberg picture operator 
$S_m^{\dagger}(\vec x_j,t) =$ $\exp(iHt)$ $S_m^{\dagger}(\vec x_j,0)$  
$\exp(-iHt)$ 
gives trivial exponentials in $t$, 
so the 
time integral  
leads to a single energy-conserving delta 
function. In isotropic antiferromagnets  
the ground state typically has $S=0$, so the accessible  
excited states $\{ |\psi_N\rangle \} $ all have $S=1$. 
We can then evaluate this exclusive structure 
factor to a specific polarization state, here $S_z=+1$, 
without loss of generality.  
For $S_z=+1$ 
the only nonzero spherical components of the exclusive structure factor is then 
\begin{displaymath} 
{\rm S}_{++}^{\psi_0\to\psi_N}(\vec k,\omega) = \hskip 5cm 
\end{displaymath} 
\begin{equation} 
\delta(E_N - E_0 - \omega) \cdot   
\bigg| 
\sum_{i}  
\langle \psi_N |  
S_+(\vec x_i) |\psi_0\rangle  
\; e^{i \vec k \cdot \vec x_i }  
\bigg|^2 
\ . 
\end{equation}  
 
If the excited state $|\psi_N\rangle$ is an eigenstate of momentum, 
which we can assume for the alternating chain without loss of 
generality, 
the matrix elements of the spin operator at translationally 
equivalent sites are equal modulo a plane wave, 
\begin{equation} 
\langle \psi_N(\vec p \, ) |  
S_m(\vec x_i) |\psi_0\rangle = 
 e^{-i \vec p \cdot (\vec x_i - \vec x_0)}  
\langle \psi_N (\vec p\, ) |  
S_m(\vec x_0) |\psi_0\rangle  
\end{equation}  
where $\vec x_0$ is some reference site.  
The sum over all  
spin sites 
$i$ can then be reduced to a sum  
over  
all sites  
in the unit cell  
$i*$  
times a momentum conserving delta function, 
\begin{displaymath} 
{\rm S}_{++}^{\psi_0\to\psi_N}(\vec k,\omega) =  
N^2_{unit\;  cells} \; 
\delta(E_N - E_0 - \omega)\,    
\delta_{\vec k, \vec p\, }  \ \cdot 
\end{displaymath} 
\begin{equation} 
\bigg| 
\sum_{sites\ i* \ in \atop unit\ cell}  
\langle \psi_N(\vec p \, ) |  
\;  
S_+(\vec x_{i*})  
\;  
|\psi_0\rangle  
\;  
e^{i \vec k \cdot \vec x_{i*} } \; 
\bigg|^2 
\ . 
\end{equation}  
 
If we are only interested in the $k$ dependence and {\it relative}  
intensities of neutron 
scattering from the  
various  
$S=1$ excitations, 
the overall normalization is irrelevant, and we can simply evaluate  
dimensionless reduced intensities. It is convenient to normalize this 
reduced exclusive structure factor as 
\begin{equation} 
{\cal S}(\vec k \,) = N_{\it unit\; cells} \; 
\bigg| 
\sum_{i*}  
\;  
\langle \psi_N(\vec k \, ) |  
S_+(\vec x_{i*})  
\; 
|\psi_0\rangle  
\;  
e^{i \vec k \cdot \vec x_{i*} }  
\;  
\bigg|^2 
\ . 
\label{strfac} 
\end{equation} 
(Here and for the remainder of the paper we suppress the 
superscript $\psi_0\to\psi_N$ on the exclusive structure factor.)
This expression can be  
evaluated analytically (using perturbation theory or other approximate 
wavefunctions) or numerically using  
wavefunctions on  
finite lattices.  
 
\subsection{One-magnon exclusive S(k)} 
 
We will now consider the reduced exclusive  
neutron scattering structure factor  
$S(k)$, 
Eq.(\ref{strfac}), 
for the excitation of  
one-magnon states in the alternating 
chain, using dimer perturbation theory and the multiprecision 
technique. 
The exclusive structure factor involves the matrix element  
 
\begin{displaymath} 
\sum_{i*=1,2}  
\langle \psi_N(\vec k) |  
S_+(\vec x_{i*})  
\; 
|\psi_0\rangle  
\; 
e^{i \vec k \cdot \vec x_{i*} }  
\;  
= 
\end{displaymath} 
\begin{equation} 
\langle \psi_1(k) |  
\ (  
S_+^L  
\; e^{-ikd/2}  
+ 
S_+^R  
\; e^{+ikd/2}  
) 
\ 
|\psi_0\rangle  \ .
\label{sf_me} 
\end{equation} 
The superscripts $L$ and $R$ refer to the left and right spins in  
the first dimer. 
Note that this can be written as a sum of 
dimer-spin conserving and changing terms, 
\begin{displaymath} 
S_+^L 
\; e^{-ikd/2}  
+ 
S_+^R 
\; e^{+ikd/2}  
= \hskip 1cm 
\end{displaymath} 
\begin{displaymath} 
\cos(kd/2)  
\underbrace{ 
\bigg( 
S_+^L 
+ 
S_+^R 
\bigg) 
}_{\rm dimer-spin\ conserving}  
\hskip 1cm 
\end{displaymath} 
\begin{equation} 
+ 
\sin(kd/2)  
\underbrace{ 
{1\over i}  
\bigg( 
S_+^L 
- 
S_+^R 
\bigg)  
}_{\rm dimer-spin\ changing}  
\ . 
\label{sf_me_2} 
\end{equation} 
 
For the one-magnon excitation 
it suffices to determine the matrix element of the raising 
operator on a single spin, because  
$S_+^L$  
and 
$S_+^R$  
have opposite one-magnon matrix elements for any $k$. (This is not true for 
a general $S=1$ excitation.) 
Evaluating  
the matrix element of $S_+^L$  
to $O(\alpha)$ using the perturbative 
$|\psi_0\rangle$ 
and  
$|\psi_1(k)\rangle$, 
Eq.(\ref{psi0_2nd}) and Eq.(\ref{psi1_1st}), we find   
\begin{equation} 
\langle \psi_1(k) |  
S_+^L  
\; 
|\psi_0\rangle  
\;  
= 
-{1\over \sqrt{L}}  
\;  
\bigg(1+{\alpha\over 4}\cos(kb) \bigg) 
\end{equation} 
so the matrix element Eq.(\ref{sf_me}) is 
\begin{displaymath} 
\sum_{i*=1,2}  
\langle \psi_N(\vec k) |  
S_+(\vec x_{i*})  
\; 
|\psi_0\rangle  
\; 
e^{i \vec k \cdot \vec x_{i*} }  
= 
\end{displaymath} 
\begin{equation} 
{1\over \sqrt{L}}  
\; 
\bigg( e^{ikd/2} - e^{-ikd/2}  \bigg) \; 
\bigg(1+{\alpha\over 4}\cos(kb) \bigg) 
\ . 
\end{equation} 
The one-magnon exclusive 
neutron scattering structure factor $S(k)$ is proportional to the 
modulus squared of this spin 
matrix element. To $O(\alpha)$ we find 
\begin{equation} 
{\cal S}(k) = 
\bigg( 1 - \cos(kd) \bigg) \; 
\bigg( 1 + {\alpha \over 2} \, \cos(kb) \bigg) 
\  . 
\label{sf_dpt}
\end{equation} 
The small-$k$ suppression $1 - \cos(kd) \propto \sin(kd/2)^2$ is  
familiar from isolated dimer problems 
and gives a basic intensity ``envelope'' 
that measures the dimer size $d$.  
(See for example Ref.\cite{our_vodpo}.) 
The separation {\it between} dimer centers 
$b$ enters as a more rapid modulation 
$1 + (\alpha/ 2)\cos(kb)$ of the  
intrinsic dimer form $1 - \cos(kd)$. This $O(\alpha)$ modulation 
arises from the excitation of the ``two-exciton'' component of 
the ground state $|\psi_0\rangle$ 
to the unperturbed ``one-exciton'' component of 
$|\psi_1(k )\rangle$, and apparently has been observed in 
recent neutron scattering experiments on 
single crystals of  
Sr$_{14}$Cu$_{24}$O$_{41}$ \cite{Nscat142441}. 
 
The ${\cal S}(k)$ series 
Eq.(\ref{sf_dpt})
may be continued using the multiprecision approach. 
We introduce a cosine expansion for 
the $S_+^L$ matrix element, 
\begin{equation} 
\langle \psi_1(k) |  
S_+^L 
\; 
|\psi_0\rangle  
= \; 
- {1\over \sqrt{L}} \; 
\sum_{\m=0}^{\infty} 
s_\m(\alpha) \, \cos(\m kb) \ , 
\end{equation} 
and the coefficients we find to $O(\alpha^3)$ are 
\begin{equation} 
\begin{array}{lrrrrr} 
s_0   
& =  
& 1  
&   
& - {11\over 64} \, \alpha^2  
& - {5\over 128}\, \alpha^3 
\\ 
\\ 
s_1  
& =  
& 
& + {1\over 4} \, \alpha  
& - {1\over 16}\, \alpha^2 
& + {31\over 1536}\, \alpha^3 
\\ 
\\ 
s_2  
& =  
& 
& 
& + {5\over 64}\, \alpha^2 
& + {31\over 384}\, \alpha^3 
\\ 
\\ 
s_3   
& =  
&   
&   
&   
& + {15\over 512}\alpha^3 
\end{array} 
\ . 
\end{equation} 
These have the same order finite size effects as the  
dispersion Fourier coefficients 
$\{ a_\m \}$,  
so diagonalization of an $L$ 
site chain gives the complete set of exclusive structure factor  
Fourier coefficients to  
$O(\alpha^{L/4})$. 
 
These coefficients determine  
the $O(\alpha^3)$ generalization of 
Eq.(\ref{sf_dpt}), which is 
\begin{displaymath} 
{\cal S}(k) = 
 \bigg( 1 - \cos(kd) \bigg) \cdot 
\end{displaymath} 
\begin{displaymath} 
\Bigg\{ 
\bigg(1 - {5\over 16}\, \alpha^2 -{3\over 32}\, \alpha^3 \bigg) 
+ 
\bigg(  {1 \over 2} \, \alpha -{1\over 8} \, \alpha^2  
- {5\over 192} \, \alpha^3 \bigg) \; \cos(kb)  
\end{displaymath} 
\begin{equation} 
+ 
\bigg(  {3\over 16 } \, \alpha^2  
+ {7\over 48} \, \alpha^3 \bigg) \; \cos(2kb)  
+ 
 {5\over 64 } \, \alpha^3  
 \; \cos(3kb)  
\Bigg\} 
\  . 
\label{sf_mp} 
\end{equation} 
 
A numerical example of this exclusive structure factor is presented in Fig.8 
for the case of (VO)$_2$P$_2$O$_7$, which will be discussed in Sec.VIII.
 
\subsection{Two-magnon bound state exclusive S(k)} 
 
To evaluate the exclusive neutron scattering 
structure factor to the $S=1$ two-magnon 
bound state we require the spin matrix element Eq.(\ref{sf_me}). This bound 
state is excited by a rather different mechanism than the one-magnon 
state discussed in Sec.VIII.B above. To leading order the bound state wavefunction is 
\begin{equation} 
|\psi_1^{(0)}(k)\rangle = 
{1\over \sqrt{N_d}} \sum_{m=1}^{N_d}\;  
e^{ikx_m} \; 
|(1,1)_{m,m+1}\rangle  
\end{equation} 
which is invisible to neutron excitation of 
the ``bare'' ground state, since two spin flips 
are required to connect these states. 
 
At 
$O(\alpha)$ a coupling to neutrons appears 
through the 
nonleading parts of the ground state and bound state. 
There are 
two such contributions, which are apparent on inspecting 
the $O(\alpha)$ states,

\begin{equation} 
|\psi_0\rangle =   
|0\rangle  
-{\sqrt{3}\over 8} \, \alpha \, 
\sum_{m=1}^{N_d} | (0,0)_{m,m+1}\rangle  
\end{equation} 
and 
 
\begin{displaymath} 
|\psi_1(k)\rangle = 
{1\over \sqrt{N_d}} \sum_{m=1}^{N_d}\;  
e^{ikx_m} \bigg[ 
|(1,1)_{m,m+1}\rangle  
\end{displaymath} 
\begin{displaymath} 
+ \ O(\alpha^0)  \  
|(1,1)_{m,n}\rangle \ {\rm terms, \ with \ } |m-n|\geq 2,  
\end{displaymath} 
\begin{equation} 
+{\alpha \over 2\sqrt{2}} \Big(1+e^{-ikb}\Big) |(+)_m\rangle + ... \bigg] \ . 
\end{equation} 
 
In the matrix element 
$ 
\langle \psi_1(k) |  
S_+(x) 
|\psi_0\rangle  
$ 
there is a $0~\to~1$ exciton coupling of the bare  
ground state to the $O(\alpha)$  
one-exciton component of the perturbed 
bound state, which has the same form as the leading one-magnon 
matrix element. 
The second contribution is the  
$2\to 2$ exciton coupling of the  
$O(\alpha)$ $|(0,0)\rangle$ two-exciton 
part of the ground state to the two-exciton bare bound state.  

These matrix elements have quite different $k$-dependences because 
the $0\to 1$ term is dimer-spin changing and the $2\to 2$ term is 
dimer-spin conserving. From 
Eq.(\ref{sf_me_2}) 
one can see that this leads to 
prefactors of $\sin(kd/2)$ and $\cos(kd/2)$ respectively. Since 
the one-magnon $S(k)$ has an overall $\sin(kd/2)$ dependence, it may  
be possible to distinguish one-magnon and  
bound state modes by the one-magnon zeros, which occur at 
at $k=2n\pi / d$. 
 
We cannot at present give a closed form result for the perturbative $S_+$  
bound state matrix elements for general $k$ 
because the $O(\alpha^0)$ mixing problem between states in the  
degenerate two-exciton manifold (discussed in Sec.VI)  
must first be solved.  
 
We also encountered difficulties in numerical studies of the bound  
state. We found very slow convergence of our projection 
method to this state, perhaps due to the presence of a nearby  
bulk-limit continuum, 
so the use of our multiprecision techniques was not practical. 
This method requires implementation of an alternative iteration scheme 
with better convergence. Work along these lines 
is in progress; we plan to present results for bound state energies and matrix elements 
in a future publication. 
 
\section{Comparison with (VO)$_2$P$_2$O$_7$} 
 
To illustrate the utility of our results we will now apply them to the 
one-magnon  
dispersion relation $\omega(k)$  
and exclusive neutron scattering intensity ${\cal S}(k)$ observed in  
(VO)$_2$P$_2$O$_7$, ``VOPO''. This material 
is an alternating chain with magnetic V$^{4+}$ ion spacings of 
3.2~\AA \ and 5.1~\AA \ along the chain pathways \cite{our_vopo}.  
The chains run along the crystallographic 
$b$ direction, and 
there is a weaker interchain coupling 
$J_a$ along the ``ladder'' direction $a$,
which gives a leading-order 
contribution of $J_a \cos(k_a)$ to $\omega(k)$.

A fit of the low-lying one-magnon 
branch $\omega([0,k_b=k,0])$ observed in VOPO to the fifth order 
formula Eq.(\ref{def_Fourier},\ref{ans_Fourier}) is shown in  
Fig.7.  
(Since VOPO has $J_a \approx -0.73$~meV
we have added  
this $J_a$ to the 
theoretical result Eq.(\ref{def_Fourier}) to obtain 
$\omega[k_a=0,k_b=k,k_c=0]$.)  
A least-squares fit gives  
an exchange of $J=11.0(1)$~meV and alternation  
of $\alpha = 0.796(4)$, which provides  
an excellent account of the data.  
We can similarly fit the more accurate ninth-order formulae 
for $E_{gap}$ and $E_{ZB}$ 
(Eqs.( \ref{Egap_mp},\ref{EZB_mp})),  
including the  
$J_a$ shift,  
to the measured values of  
3.1(1)~meV and 15.4(3)~meV;  
this gives consistent values of  
$J=10.92$~meV and $\alpha=0.798$.  
 
The intensity of the  
modes seen in neutron scattering 
is quite sensitive to the spatial geometry of the  
alternating chain. For VOPO we can use this relation 
to deduce which is the stronger of the two inequivalent exchange paths 
in the chain.  
Fig.8 shows the theoretical one-magnon  
intensity variation over a wide range of $k$.
The dashed line is the 
$(1-\cos(kd))$ scattering intensity for isolated dimers, and the solid 
line shows 
the $O(\alpha^3)$ alternating-chain result 
Eq.(\ref{sf_me}) 
for ${\cal S}(k)$, 
with $\alpha=0.8$ and $d/b = 5.1$~\AA / $8.3$~\AA . 
This $d/b$ ratio assumes that the long 
V-PO$_4$-V dimer, which has $d=5.1$~\AA, has the stronger interaction $J$.
The characteristic variation of ${\cal S}(k)$, which is measured
in neutron scattering, allows 
a direct check of this bond assignment and is a stringent test of the 
alternating chain model as applied to
VOPO 
(Ref.\cite{new_vopo}, in preparation).

\section{Summary and conclusions}

In summary, we have presented many new  
analytical and numerical results for the low-lying 
excitations of the spin-1/2 alternating Heisenberg chain. We introduced 
the model in Sec.I and II, and Sec.III  
reviewed previous work. 
Sec.IV.A  
introduced perturbation theory about the dimer limit,
and
in Sec.IV.B
we used this approach to 
confirm previous results for the  
one-magnon dispersion $\omega (k)$ to third order, and  
extended the ground state energy calculations to fourth  
order. 
Sec.IV.C summarized the expected critical behavior of the
gap and ground-state energy as we approach the uniform chain limit. 
Relations between energies and their derivatives
at the critical point were also derived.  
To complement these analytical calculations, in Sec.V.A  
we undertook numerical calculations of 
bulk limit energies of the ground state, gap, zone boundary, 
and $S=0$ and $S=1$ two magnon bound states, using Lanczos methods on systems 
up to $L=28$.  
As a major part of this work, in Sec.V.B we  
introduced a novel technique  
in which multiple precision 
calculations with a very small dimer coupling are used  
to infer perturbation expansions to high order.  
This confirmed third and fourth order energy 
expansions and allowed continuation of the $e_0$, $E_{gap}$ and 
$E_{ZB}$ series to $O(\alpha^9)$. 
The one-magnon dispersion relation 
$\omega (k)$ was then derived to $O(\alpha^5)$ using this method. 
The expected critical behavior of the ground state energy and energy gap  
was compared with our numerical results
in Sec.V.C. 
The ground state energy was found to be in approximate agreement  
with the expected (to within logarithmic corrections) 
$\delta^{4/3}$ power law.
The singlet-triplet gap however 
appears to support a
power law of $\delta^{3/4}$ rather than $\delta^{2/3}$ over the accessible
range of $\delta$.
Sec.VI considered two-magnon bound states. 
At the zone boundary $k=\pi/b$ an attractive potential  
between magnons was found to give rise to $S=0$ and $S=1$  
two-magnon bound states for all alternations. Numerical results and 
$O(\alpha^2)$ analytical forms  
were given 
for the magnon-magnon binding  
energy.  
The $S=0$ bound state 
was found for all $k$ considered, and numerical results for its 
binding energy at $k=0$ were also given.  
The $S=1$ bound state however was found to 
exist only for a range of $k$ near $\pi/b$.
In Sec.VII we introduced an exclusive neutron scattering structure 
factor ${\cal S}(k)$; this was evaluated  
analytically for the one-magnon band 
using the multiple precision technique and dimer perturbation theory. 
The one-magnon exclusive structure factor 
was determined  to $O(\alpha^3)$, and was presented as a modulation times 
the familiar isolated-dimer form $(1-\cos(kd))$. 
Finally, in Sec.VIII 
we gave an illustrative 
application of these results to recent neutron scattering data 
on \vopons,
which is dominantly an alternating chain. We showed that the 
fifth-order dispersion relation 
gives an excellent fit to the observed one-magnon band, and noted that 
the predicted exclusive 
structure factor ${\cal S}(k)$ will allow a detailed test of the
alternating chain model as applied to 
\vopons.
 
In conclusion, we have used perturbation theory and  
numerical methods to calculate the properties of 
the ground state and one- 
and two-magnon states in the alternating Heisenberg chain. 
Using a new technique based on multiple precision 
programming we have derived high-order series expansions 
for energies and matrix elements in the alternating chain. 
This technique is quite general and should be applicable to  
many other quasi-1D quantum spin systems. 
 
\section{Acknowledgements} 
 
We acknowledge useful discussions with  
D.Bailey, E.Dagotto, R.S.Eccleston, V.Emery
A.W.Garrett,  
S.E.Nagler, D.J.Scalapino, H.J.Schulz, P.D.Stevenson and M.R.Strayer. 
 
This work was supported in part by the United States Department 
of Energy under contract  
DE-AC05-96OR22464 at the 
Oak Ridge National Laboratory, managed for the U.S. D.O.E. 
by Lockheed Martin Energy Research Corporation  
 
\newpage 
 
\begin{center} 
\begin{table} 
\caption{Representative alternating chain materials. 
The parameters ${\cal J} = (1+\alpha)J/2$ and $\delta=(1-\alpha)/(1+\alpha)$ 
are also commonly used (see text).} 
\begin{tabular}{lccc} 
Material  & $J$(meV) &  $\alpha$ & reference \\ 
\tableline 
Sr$_{14}$Cu$_{24}$O$_{41}$ & 11.2 & $-$0.10(1) & \cite{Nscat142441,ladderite}  \\ 
Cu(NO$_3$)$_2$ $\cdot 2.5$ H$_2$O 
  & 0.45  & 0.27  & 
\cite{DBGKP,cuno3} \\ 
\vopo & $\approx 10 $ & 0.8 & \cite{our_vopo} \\ 
CuWO$_4$ & $\approx$ 12 & $\approx$ 0.9   & \cite{lake} \\ 
\navo & 40  & 0.9 & \cite{navo}  \\ 
CuGeO$_3$\footnote{There may also  
be important second nearest neighbor interactions  
in this material.} & 13 & 0.94  & \cite{CCE} \\ 
\end{tabular} 
\end{table} 
\end{center} 
 
\onecolumn 
 
\begin{center} 
\begin{table} 
\caption{Bulk limit alternating chain energies ($J=1$) and two-magnon
bound state binding energies, Eq.(44),  
extrapolated from $L=20,24,28$. The change observed in going from an 
$L=16,20,24$  
extrapolation to $L=20,24,28$ (an estimate of the systematic error)  
is given in parenthesis.} 
\begin{tabular}{cllllll} 
 $\alpha $   &  
\ \ \ \ \ $E_0/L$   &  
$\omega(k=0)=E_{gap}$   &  
$\omega(k=\pi/b)$   &  
$E^{S=0}_B(k=0)$   &  
$E^{S=0}_B(k=\pi/b)$   &  
$E^{S=1}_B(k=\pi/b)$    \\ 
\tableline 
0.0 [exact]  
    &  -0.375 = -3/8   &  1              &  1   
& 0 & 0  &  0 \\ 
0.1 &  -0.375480805    &  0.946279339    & 1.051248884    
& 0.0002   & 0.0456   & 0.0210    \\ 
0.2 &  -0.376974494    &  0.885209996    & 1.104980718    
& 0.0009(-2)  & 0.0824   & 0.0343    \\ 
0.3 &  -0.379566321    &  0.816844275(+1)    &  1.161143536   
& 0.0025(-1)   & 0.1104   & 0.0403    \\ 
0.4 &  -0.383356250    &  0.74106141(+3) & 1.219628893(+1)    
& 0.0049(-3)   & 0.1294   & 0.0402   \\ 
0.5 &  -0.388465614    &  0.6574777(+5)  & 1.280237618(+9)    
& 0.0077(+34)   & 0.1394   & 0.0353   \\ 
0.6 &  -0.395048423(-3) &  0.565296(+7)  & 1.3426173(+2)    
&    & 0.1405   & 0.0280(+1)  \\ 
0.7 &  -0.40331243(-5)  &  0.46298(+5)   & 1.406138(+3)    
&    & 0.1327(-1)   &  0.0207(+17) \\ 
0.8 &  -0.4135644(-8)   &  0.3474(+3)    & 1.46959(+4)    
&    & 0.1151(-1)   &   \\ 
0.9 &  -0.426330(-16)   &  0.2098(+17)   & 1.5298(+8)    
&    & 0.0829(+2)   &   \\ 
1.0  & -0.44314718...  &  0  &  1.5707963...        
&    &  0  &   0    \\ 
     &  $= 1/4 - \ln(2)$ &    & =$\pi/2$         
&    &    &      \\ 
\tableline

\end{tabular} 
\end{table} 
\end{center}

\newpage 
 
\begin{center} 
{\Large Figure Captions} 
\end{center}

\begin{figure} 
{Figure~1. 
The geometry of a 1D alternating chain. The internal dimer  
spin-spin interaction (solid line)  
is $J_1=J$ and the dimer extent is $d$.  The  
spin-spin coupling between dimers (dashed line) is $J_2=\alpha J$, and the 
spacing between dimer centers, which is the length of the  
unit cell, is $b$. A spatially uniform chain has 
a smaller unit cell length of $d = b/2$, which is normally called $a$.}  
\end{figure} 
 
\begin{figure} 
{Figure~2. 
Ground state energy per spin $e_0(\alpha) = E_0/LJ$ 
of the alternating chain. The dashed line 
is third-order perturbation theory, the solid line is ninth order, and the points are bulk limit 
extrapolations of Lanczos data.} 
\end{figure} 
 
 \begin{figure} 
{Figure~3. 
Singlet-triplet energy gap $E_{gap}/J$ of the alternating chain, as in Fig.2.} 
\end{figure} 
 
\begin{figure} 
{Figure~4. 
Dispersion  $\omega(k)$ of the one-magnon band in the alternating chain  
for  
$\alpha=0.2,0.4,0.6,0.8$ and $1.0$,  
using the fifth-order dispersion relation 
Eqs.(\ref{def_Fourier},\ref{ans_Fourier}). 
The $\alpha=1$ curve is the exact result $\pi  |\sin(kb/2)| / 2$. } 
\end{figure} 
 
\begin{figure} 
{Figure~5. 
Critical behavior of the alternating chain as a function of $\delta$, 
with ${\cal J}$ fixed.  
Fig.5a shows the energy per spin relative to the bulk limit,  
$\tilde e_0(\delta) - \tilde e_0(0)$, and a fit to
$c_1 \delta^{4/3}$ which gives $c_1=0.3134$.
Fig.5b shows $E_{gap}/{\cal J}$ and a fit to $c_1\delta^{3/4}$
which gives $c_1=2.003$. 
} 
\end{figure} 
 
\begin{figure} 
{Figure~6. 
Binding energies of the $S=0$ and $S=1$ two-magnon bound states at $k=\pi / b$ 
(band maximum). The points are bulk-limit extrapolations of Lanczos data (Table II) and the lines are second order perturbation theory, Eq.(\ref{EB_mp}).} 
\end{figure} 
 
\begin{figure} 
{Figure~7. 
A fit of the alternating chain dispersion relation 
Eqs.(\ref{def_Fourier},\ref{ans_Fourier}) to 
the one-magnon dispersion observed in neutron scattering from
(VO)$_2$P$_2$O$_7$ 
\cite{our_vopo}, as discussed in Sec.VII.
The fitted parameters are $J=11.0$~meV and $\alpha=0.796$. 
}  
\end{figure}

\begin{figure} 
{Figure~8. 
The predicted neutron scattering intensity (exclusive structure factor) 
${\cal S}(k)$ 
for the one-magnon mode 
in (VO)$_2$P$_2$O$_7$, Eq.(\ref{sf_mp}).  
The parameters are 
$\alpha=0.8$ (from a fit to the 
dispersion), $d=5.1$ \AA \ and $b=8.3$ \AA. 
The dashed line shows the isolated dimer form $(1-\cos(kd))$ for comparison. 
}  
\end{figure} 
 

\begin{references} 
 
\bibitem{NSM} For early work in this area see  
P.L.Nordio, Z.G.Soos and H.M.McConnell, Ann. Rev. Phys. Chem. 17, 
237 (1966). 
 
\bibitem{CGO_Nscat} M.Nishi, O.Fujita and J.Akimitsu, Phys. Rev. B50, 
6508 (1994). 
 
\bibitem{CGO_2D} M.A\"{\i}n {\it et al.} Phys. Rev. Lett. 78, 1560 (1997).  
 
\bibitem{our_vopo} A.W.Garrett, S.E.Nagler, D.A.Tennant, B.C.Sales and 
T.Barnes,  
Phys. Rev. Lett. 79, 745 (1997). 
 
\bibitem{our_vopowder} A.W.Garrett, S.E.Nagler, T.Barnes and B.C.Sales, 
Phys. Rev. B55, 3631 (1997). 
 
\bibitem{DB} W.Duffy and K.P.Barr, Phys. Rev. 165, 647 (1968). 
 
\bibitem{BB} 
J.Bonner and H.W.J.Bl\"ote,  
Phys. Rev. B25, 6959 (1982). 
 
\bibitem{N} M.P.M.den Nijs, Physica 95A, 449 (1979). 
 
\bibitem{CF}  
M.C.Cross and D.Fisher, Phys. Rev. B19, 402 (1979). 
 
\bibitem{BE}  
J.L.Black and V.J.Emery, Phys. Rev. B23, 429 (1981). 
 
\bibitem{SKM} Z.G.Soos, S.Kuwajima and J.E.Mihalick, Phys. Rev. B32, 3124 
(1985). 
 
\bibitem{SFL} G.Spronken, B.Fourcade and Y.L\'epine, Phys. Rev. B33, 1886 
(1986). 
 
\bibitem{US}  G.S.Uhrig and H.J.Schulz, Phys. Rev. B54, R9624 (1996). 
 
\bibitem{BKJ}  G.Bouzerar, A.P.Kampf and G.I.Japaridze, 
``Elementary Excitations in Dimerized and Frustrated Heisenberg Chains'', 
cond-mat/9801046 (Jan.1998). 
 
\bibitem{FJ}  
A.Fledderjohann and C.Gros,  
Euro. Phys. Lett. 37, 189 (1997) 
(cond-mat/9612013). 

\bibitem{PRHBH}
D. Poilblanc, J. Riera, C. A. Hayward, C. Berthier and M.
Horvati\`c,  Phys. Rev. B 55, 11941 (1997).
 
\bibitem{DBGKP} 
K.M.Diedrix, H.W.J.Bl\"ote, J.P.Groen, T.O.Klassen, and N.J.Poulis, 
Phys. Rev. B19, 420 (1979). 
 
\bibitem{BR} T.Barnes and J.Riera, Phys. Rev. B50, 6817 (1994). 
 
\bibitem{BH} A.Brooks Harris, Phys. Rev. B7, 3166 (1973). 
 
\bibitem{josesref} B.N.Parlett, ``The Symmetric Eigenvalue Problem''
(Prentice-Hall, Englewood Cliffs, New Jersey, 1980).
 
\bibitem{MPFUN} D.Bailey, ``Multiprecision Translation and Execution of 
Fortran Programs'', ACM Transactions on Mathematical Software, 19,  
288 
(1993).  
 
\bibitem{mLmethod} 
E. Dagotto and A. Moreo, Phys. Rev. D31, 865 (1985);  
E. Gagliano, E. Dagotto, A. Moreo, 
and F. Alcaraz, 
Phys. Rev. B34, 1677 (1986), Erratum B35, 8562 (1987). 
 
\bibitem{SteveW} S.White, 
Phys. Rev. Lett. 69, 2863 (1993);
Phys. Rev. B48, 10345 (1993).
 
\bibitem{our_vodpo}  
D.A.Tennant,  
S.E.Nagler,  
A.W.Garrett,  
T.Barnes  and C.C.Torardi, 
Phys. Rev. Lett. 78, 4998 (1997). 
 
\bibitem{Nscat142441} R.S.Eccleston,  
M.Uehara, J.Akimitsu, H.Eisaki, N.Motoyama and S.Uchida, 
cond-mat/9711053. 
See also 
M.Matsuda and  K.Katsumata,  
Phys. Rev. B53, 12201 (1996); 
M.Matsuda, K.Katsumata, H.Eisaki, N.Motoyama, S.Uchida, S.M.Shapiro and 
G.Shirane, Phys. Rev. B54, 12199 (1996); 
M.Matsuda, K.Katsumata,   
T.Yokoo, S.M.Shapiro and   
G.Shirane, Phys. Rev. B54, R15626 (1996); 
M.Matsuda, K.Katsumata, T.Osafune, N.Motoyama, H.Eisaki, S.Uchida,  
T.Yokoo, S.M.Shapiro,  
G.Shirane and J.L.Zarestky, Phys. Rev. B56, 1 (1997). 
 
\bibitem{cuno3}  
K.M.Diedrix, J.P.Groen, L.S.J.M.Henkens, T.O.Klassen, N.J.Poulis, 
Physica 93B, 99 (1978); 
D.B.Brown, J.A.Donner, J.W.Hall, S.R.Wilson, D.J.Hodgson and W.E.Hatfield, 
Inorg. Chem. 18, 2635 (1979);  
J.C.Bonner, S.A.Friedberg, H.Kobayashi, D.L.Meier and H.W.J.Bl\"ote, 
Phys. Rev. B27, 248 (1983). 
 
\bibitem{CCE} G.Castilla, S.Chakravarty and V.J.Emery,  
Phys. Rev. Lett. 75, 1823 (1995). CuGeO$_3$ appears to have a large second 
nearest neighbor magnetic coupling; see also J.Riera and A.Dobry,  
Phys. Rev. B51, 16098 (1995). 
 
\bibitem{lake}  
B.Lake, ``Neutron Scattering Studies of Alternating Chain Antiferromagnets'', 
Oxford University thesis (1997); 
B.Lake, R.A.Cowley and D.A.Tennant, 
J. Phys.: Condens. Matter 9, 10951 (1997); 
J.P.Doumerc, J.M.Dance, J.P.Chaminade, M.Pouchard, P.Hagemuller and  
M.Krussanova, Mat. Res. Bull. 16, 985 (1981). 
 
\bibitem{navo} M.Isobe and Y.Ueda, J. Phys. Soc. Jpn. 65, 1178 (1996);
D.Augier, D.Poilblanc, S.Haas, A.Delia and E.Dagotto, 
Phys. Rev. B56, R5732 (1997) 
(cond-mat/9704015). 
 
\bibitem{ladderite} E.M.McCarron III, M.A.Subramanian, J.C.Calabrese  
and R.L.Harlow, 
Mater. Res. Bull. 23, 1355 (1988);  
T.Siegrist, L.F.Schneemeyer, S.A.Sunshine, J.V.Waszczak and R.S.Roth, 
Mater. Res. Bull. 23, 1429 (1988). 
 
\bibitem{new_vopo} A.W.Garrett, S.E.Nagler, D.A.Tennant, B.C.Sales and 
T.Barnes, in preparation. 
 
\end{references}
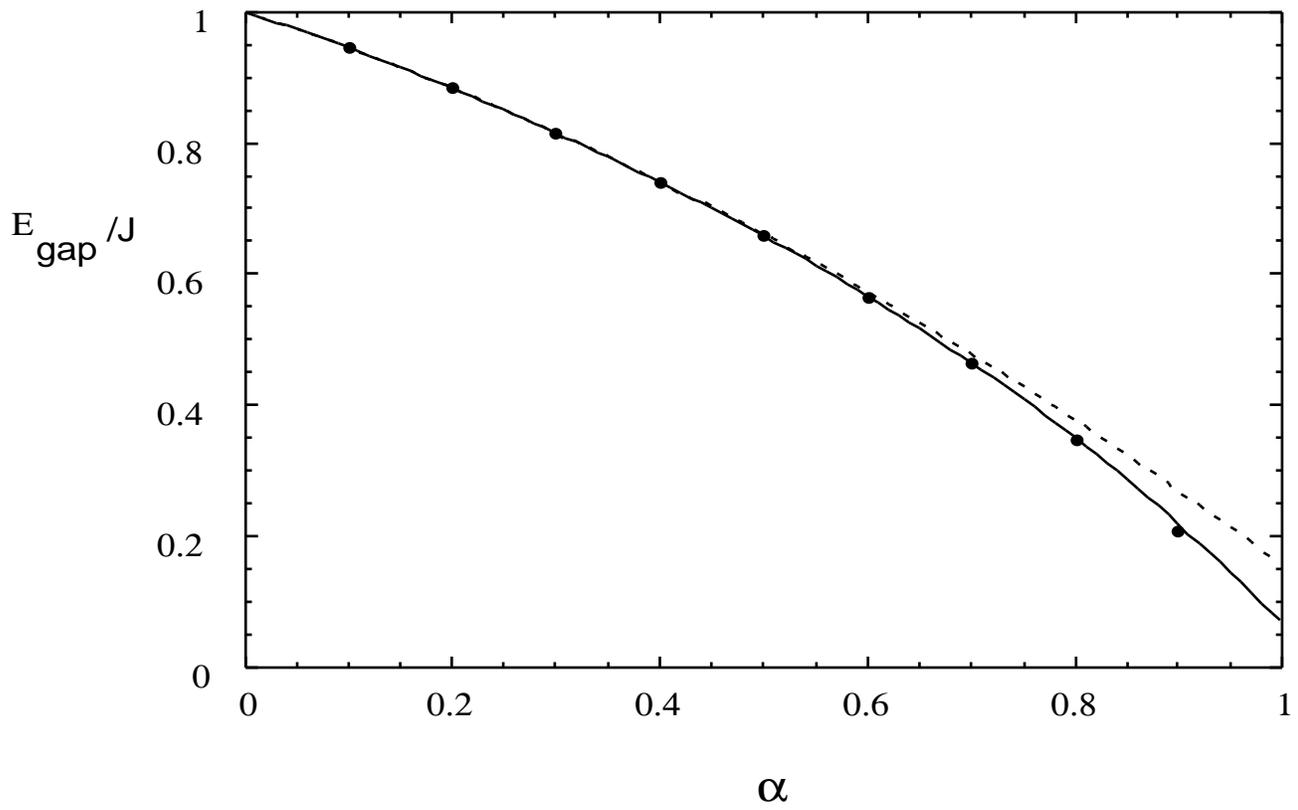
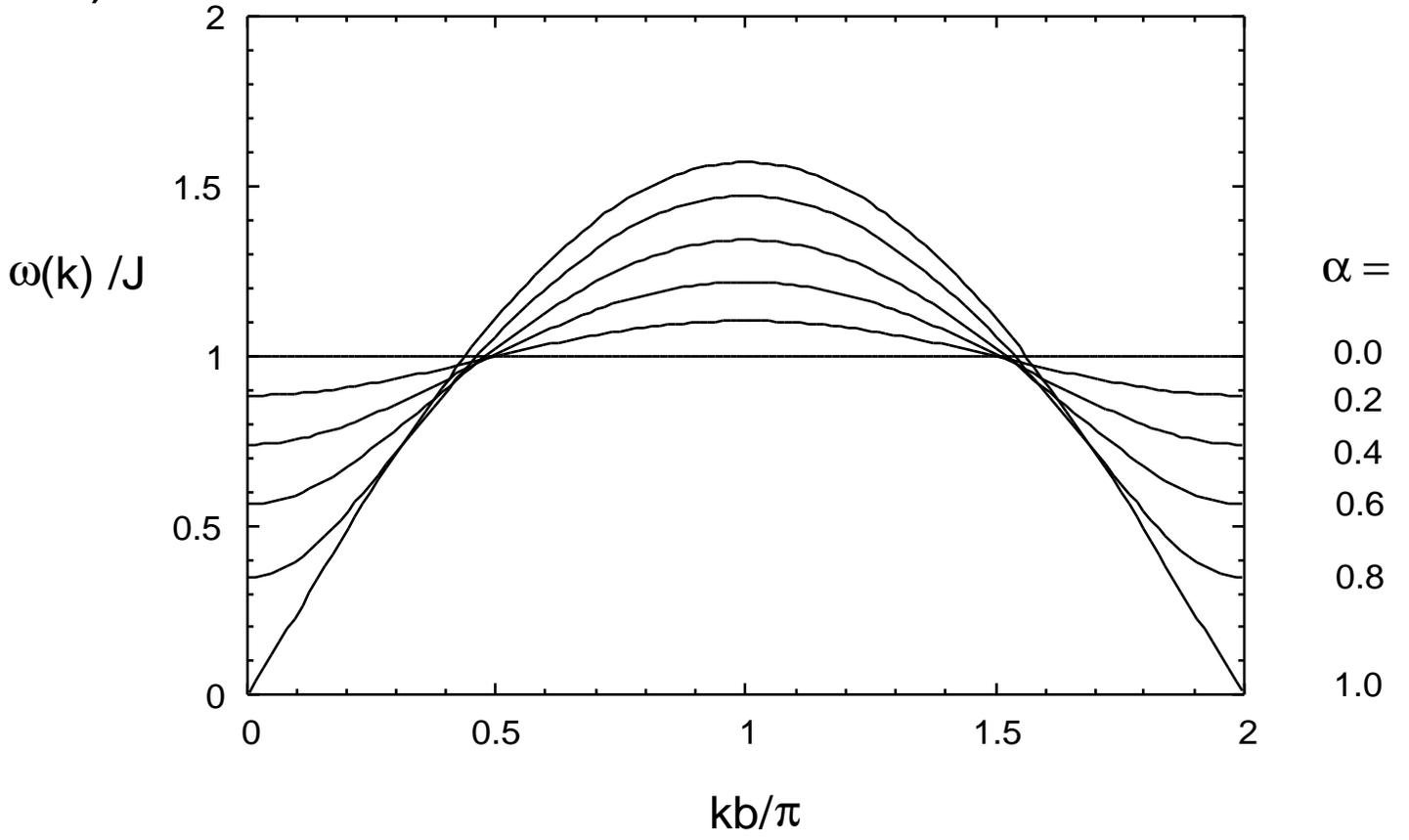
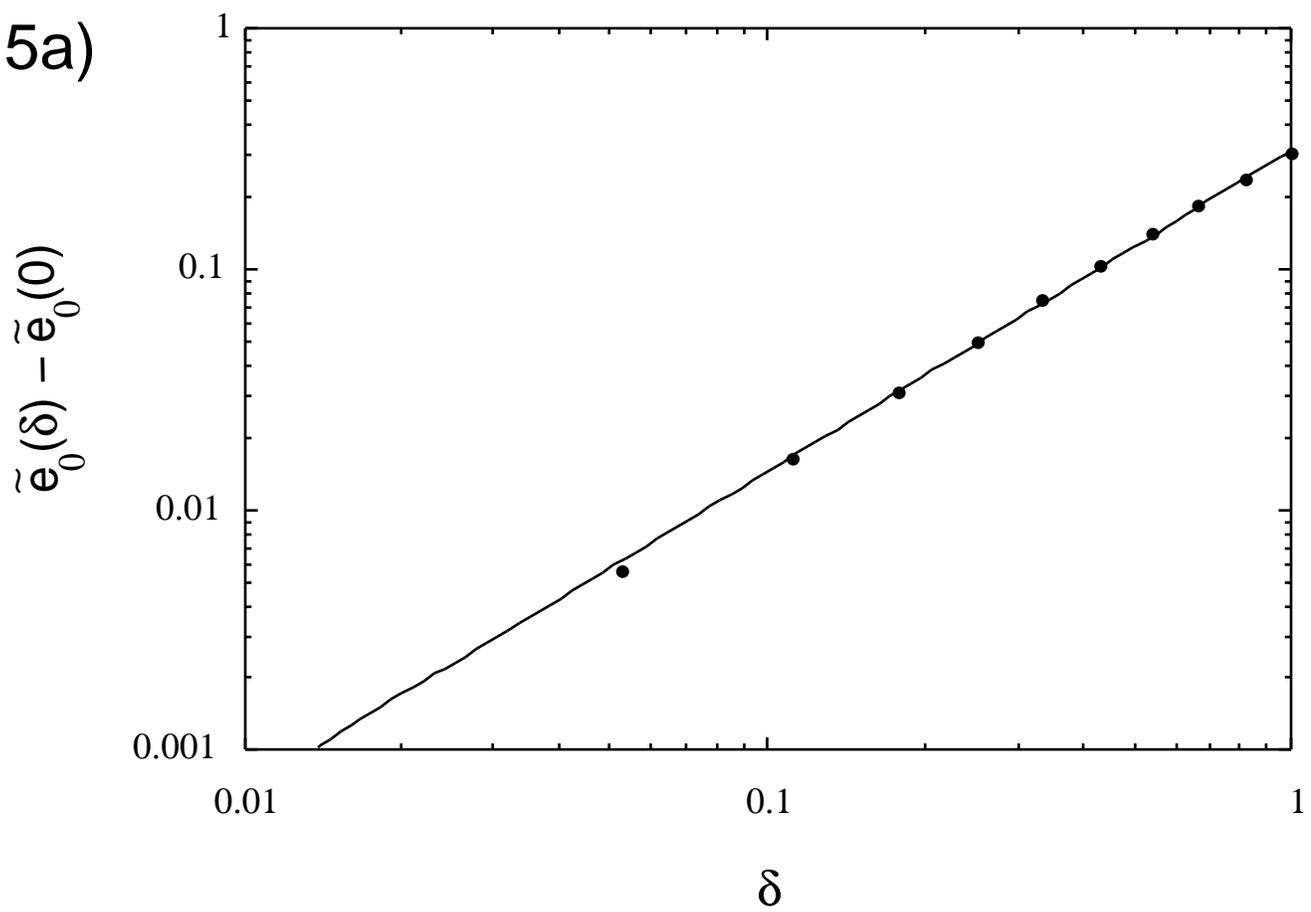
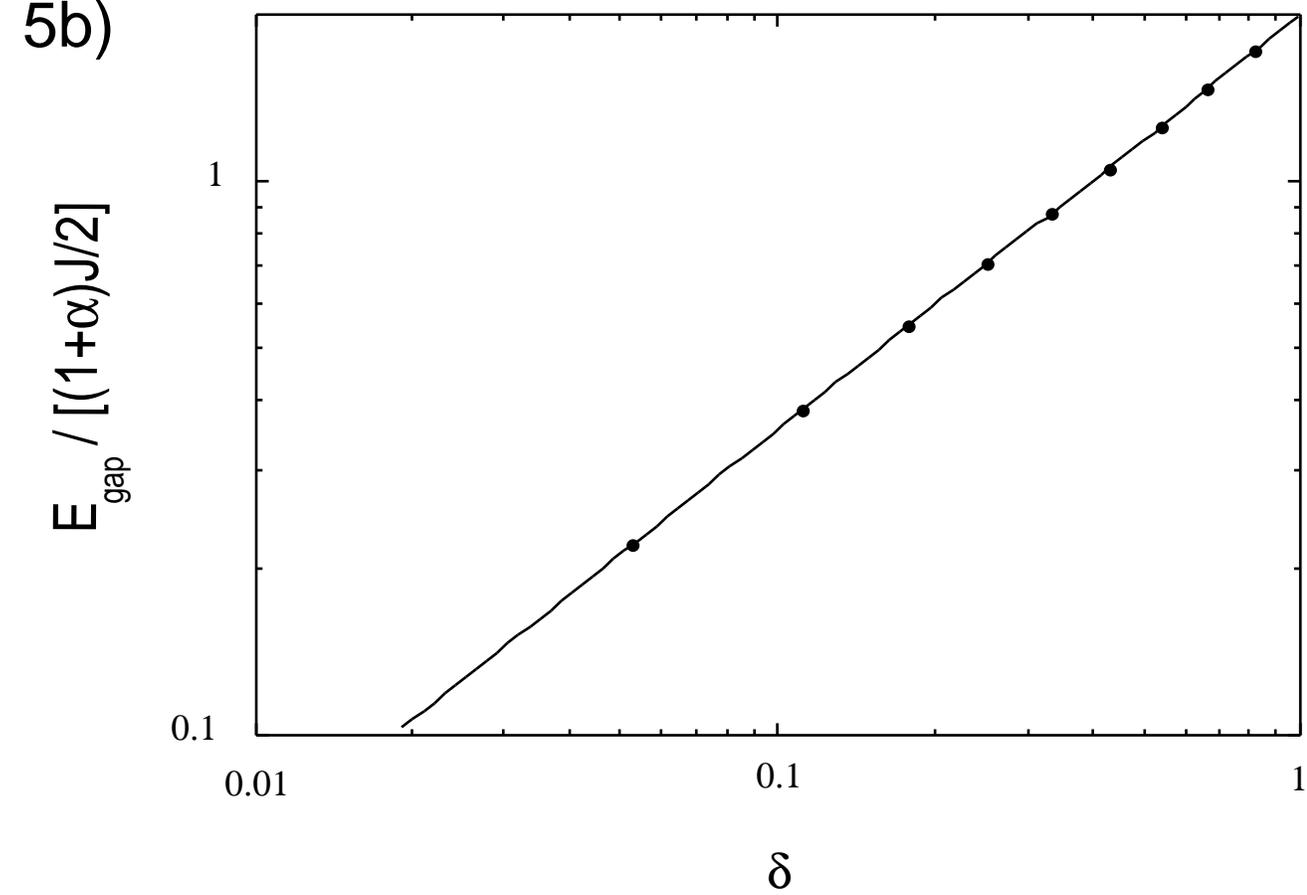
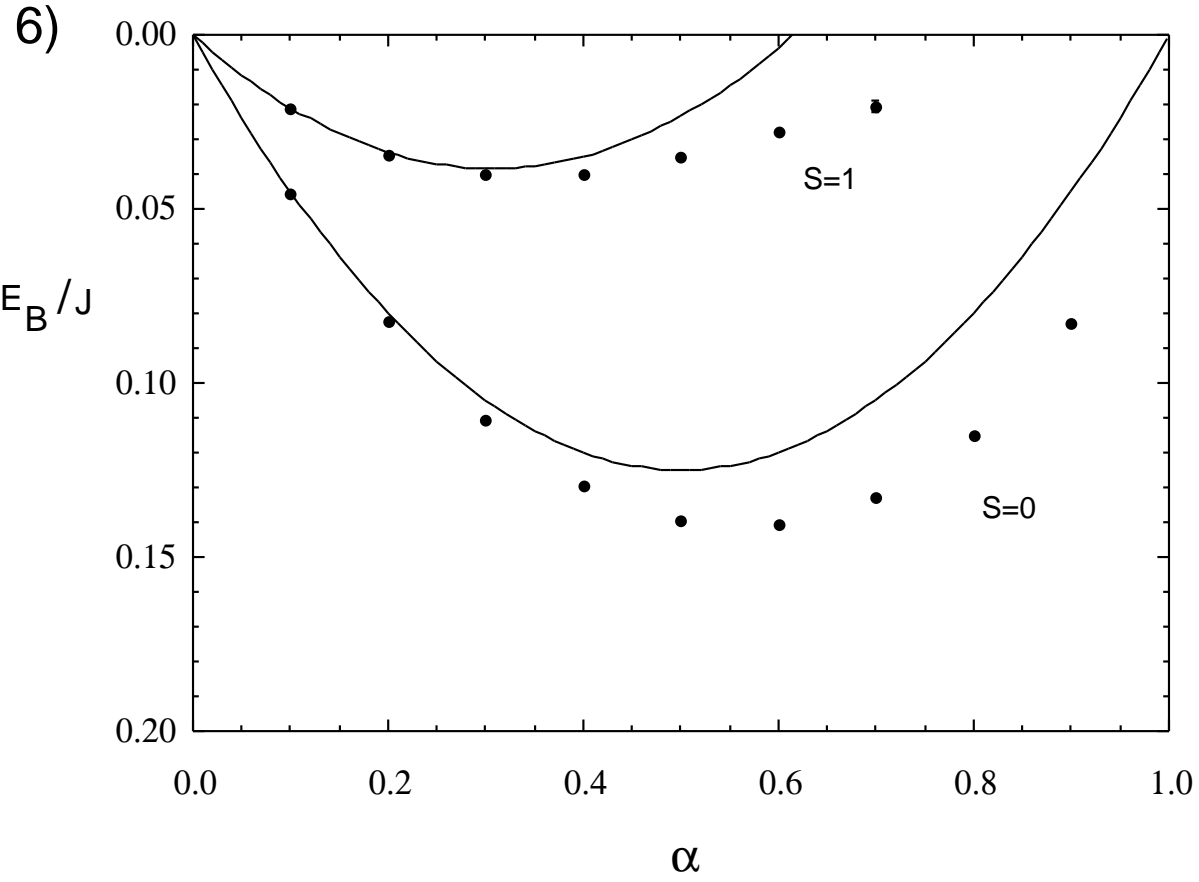
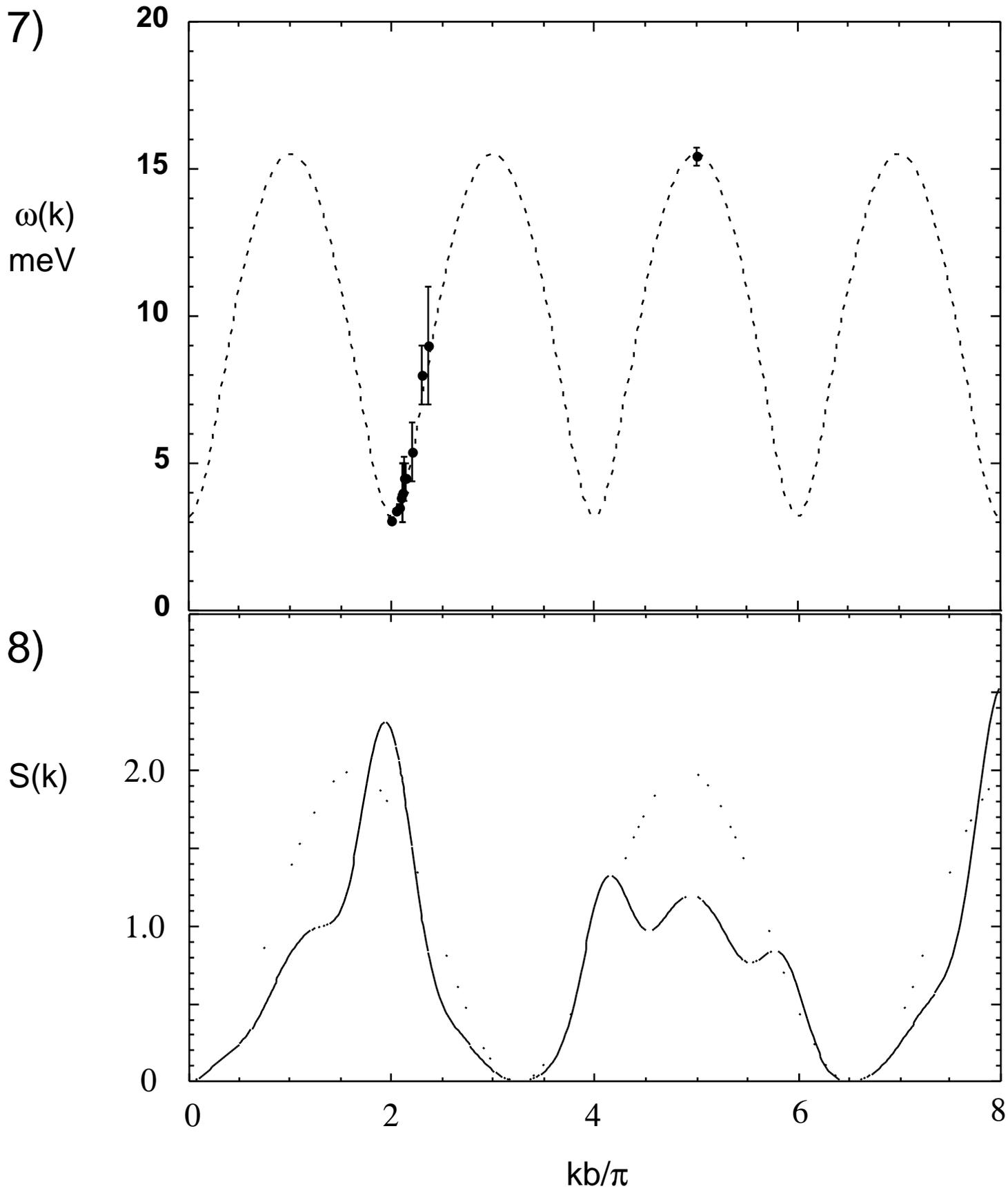
\end{document}